\documentclass[12pt,preprint]{aastex}

\newcommand{\dmm}{\mbox{$\Delta$m$_{15}(B)$}}
\newcommand{\kms}{\mbox{km s$^{-1}$}}

\shorttitle{SN 2003gs}
\shortauthors{Krisciunas et al.}

\begin{document}

\title{The fast declining Type Ia supernova 2003gs, and evidence for a significant dispersion
in near-infrared absolute magnitudes of fast decliners at maximum light\altaffilmark{1}}

\author{Kevin Krisciunas,\altaffilmark{2,3}
G. H. Marion,\altaffilmark{2,4}
Nicholas B. Suntzeff,\altaffilmark{2,3}
Guillaume Blanc,\altaffilmark{5}
Filomena Bufano,\altaffilmark{6}
Pablo Candia,\altaffilmark{7}
Regis Cartier,\altaffilmark{8}
Nancy Elias-Rosa,\altaffilmark{9}
Juan Espinoza,\altaffilmark{10}
David Gonzalez,\altaffilmark{10}
Luis Gonzalez,\altaffilmark{11}
Sergio Gonzalez,\altaffilmark{11}
Samuel D. Gooding,\altaffilmark{2}
Mario Hamuy,\altaffilmark{8}
Ethan A. Knox,\altaffilmark{12}
Peter A. Milne,\altaffilmark{13}
Nidia Morrell,\altaffilmark{11}
Mark M. Phillips,\altaffilmark{11}
Maximilian Stritzinger,\altaffilmark{11}
and Joanna Thomas-Osip\altaffilmark{11}
}

\altaffiltext{1}{Based in part on observations taken at the Cerro Tololo 
Inter-American Observatory, National Optical Astronomy Observatory, which is 
operated by the Association of Universities for Research in Astronomy, Inc. (AURA) 
under cooperative agreement with the National Science Foundation.  The near-IR
photometry from La Silla and Paranal was obtained by the European Supernova
Collaboration (ESC).}

\altaffiltext{2}{Texas A\&M University, Department of Physics, 4242 TAMU,
College Station, TX 77845-4242; {krisciunas@physics.tamu.edu},
{sam.gooding86@gmail.com}, {suntzeff@physics.tamu.edu} }

\altaffiltext{3}{George P. and Cynthia Woods Mitchell Institute for
Fundamental Physics \& Astronomy, 
Texas A\&M University, Department of Physics, 4242 TAMU,
College Station, TX 77845-4242}

\altaffiltext{4}{University of Texas, Department of Astronomy, Austin, TX 78712; 
{hman@astro.as.utexas.edu}}

\altaffiltext{5}{APC, UMR 7164, CNRS, Universit\'{e} Paris 7, CEA, Observatoire de Paris,
11 place Marcelin Berthelot, F-75231, Paris, France; {blanc@apc.univ-paris7.fr}}

\altaffiltext{6}{INAF, Osservatorio Astronomico di Padova, Vicolo dell'Osservatorio 5, 
I-35122, Padova, Italy; {filomena.bufano@oapd.inaf.it}}

\altaffiltext{7}{AURA/Gemini Observatory, Casilla 603, La Serena, Chile;
  {pcandia@gemini.edu}}

\altaffiltext{8}{Universidad de Chile, Departamento de Astronom\'{i}a, Casilla 36-D,
Santiago, Chile;  {rcartier@das.uchile.cl}, {mhamuy@das.uchile.cl}}

\altaffiltext{9}{Spitzer Science Center, California Institute of Technology, 1200 E. 
California Blvd., Pasadena, CA 91125; {nelias@ipac.caltech.edu}}

\altaffiltext{10}{Cerro Tololo Inter-American Observatory, Casilla
  603, La Serena, Chile; {juan@ctio.noao.edu}}

\altaffiltext{11}{Las Campanas Observatory, Casilla 601, La Serena, Chile;
  {nmorrell@lco.cl} {mmp@lco.cl} {mstritzinger@lco.cl} {jet@lco.cl}}

\altaffiltext{12}{Humboldt State University, 1 Harpst Street, Arcata, CA 95521; 
{eaknox@gmail.com}}

\altaffiltext{13}{University of Arizona, Steward Observatory, 933 N. Cherry Ave., 
Tucson, AZ 85719;  {pmilne511@cox.net}}

\begin{abstract}
  
We obtained optical photometry of SN~2003gs on 49 nights, from 2 to
494 days after T($B_{max}$).  We also obtained near-IR photometry on 21 nights.  SN~2003gs 
was the first fast declining Type Ia SN that has been well
observed since SN~1999by.  While it was subluminous in optical bands compared to more
slowly declining Type Ia SNe, it was not subluminous at maximum light in the near-IR
bands. There appears to be a bimodal distribution in the near-IR absolute magnitudes of
Type Ia SNe at maximum light.  Those that peak in the near-IR after T($B_{max}$) are
subluminous in the all bands.  Those that peak in the near-IR prior to T($B_{max}$),
such as SN~2003gs, have effectively the same near-IR absolute magnitudes at maximum 
light regardless of the decline rate \dmm. 


Near-IR spectral evidence suggests that opacities in the outer layers of SN~2003gs are 
reduced much earlier than for normal Type Ia SNe.  That may allow $\gamma$ rays that 
power the luminosity to escape more rapidly and accelerate the decline rate.  This 
conclusion is consistent with the photometric behavior of SN~2003gs in the IR, which 
indicates a faster than normal decline from approximately normal peak brightness.


 \end{abstract} 

\section{Introduction}  

Type Ia supernovae (SNe) have been used to
obtain accurate cosmological distances, leading to the conclusion that the
expansion of the universe is accelerating \citep{Rie_etal98, Per_etal99}.
Several pressing questions regarding the observed properties and the
underlying physics of the explosion remain unanswered.

The remarkable homogeneity of Type Ia SNe has led to general consensus
regarding the nature of the progenitor system.  Type Ia SNe are believed to
result from the thermonuclear disruption of a carbon-oxygen white dwarf. In
the favored scenario, the white dwarf is one component of a close binary, and
accretes hydrogen from its companion star. When the white dwarf reaches the
Chandrasekhar limit, explosive carbon burning sets in. Burning to nuclear
statistical equilibrium ensues, yielding radioactive $^{56}$Ni.

Although the intrinsic peak brightness of Type Ia SNe can vary by up to 2.5
magnitudes in the $B$-band, there exists a tight correlation between the peak
luminosity and the shape of light curve which can be exploited to derive
accurate relative distances. The peak brightness is determined by the amount
of radioactive $^{56}$Ni produced in the explosion. This amount spans 0.09 --
0.93$M_\odot$, although most events cluster in the 0.4-0.7$M_\odot$ range
\citep{stritzinger:06}. The cause of this large variation is not understood.
It is becoming increasingly clear that Type Ia SNe are not a one-parameter
family \citep{benetti:04b}.

No two events illustrate this more than the intrinsically subluminous
SN~1991bg \citep{filippenko:92b,leibundgut:93} and the intrinsically
overluminous SN~1991T \citep[e.g.][]{filippenko:92a}, both of which showed
photometric and spectroscopic behavior that deviated substantially from
normal Type Ia SNe, although other highly peculiar events have been discovered
recently, such as SN~2002cx \citep{Li_etal03, Bra_etal04, Jha_etal06}.

One way of gaining physical insight is to assemble a sample of Type Ia SNe
with complete observational coverage from pre-maximum to nebular phases, over
as wide a wavelength range as possible. Such data can then be used to confront
state-of-the-art explosion and spectral synthesis models. Until recently, such
data have been non-existent. For the very nearby events ($v_{helio} \la
3000$\,\kms), this shortcoming resulted in the setting up of the European
Supernova Collaboration (ESC)
\footnote[14]{http://www.mpa-garching.mpg.de/\textasciitilde rtn/} which has
already provided high quality data for at least half a dozen Type Ia SNe:
2002bo \citep{benetti:04a},
2002cv \citep{Eli_etal08},
2002dj \citep{Pig_etal08},
2002er \citep{pignata:04,kotak:05},
2003cg \citep{eliasrosa:06},
2003du \citep{stanishev:07},
2004dt \citep{Alt_etal07},
2004eo \citep{pastorello:07b}, 
2005bl \citep{Tau_etal08}, and
2005cf \citep{garavini:07,pastorello:07a}.

Important observational initiatives have been carried out at Las Campanas Observatory
and Cerro Tololo Iner-American Observatory whose goal was to obtain well-sampled
optical and near-IR light curves.  These endeavors have not been limited to nearby
Type Ia SNe.

Since \citet{Psk77}, \citet{Phi93}, and \citet{Ham_etal95} showed
that the absolute magnitudes at maximum light of Type Ia supernovae were correlated
with their decline rates,\footnote[15]{The light curve decline parameter, \dmm, is
defined as the decline in apparent brightness in the first 15 days following
$B$-band maximum. The observed range of \dmm\ is 0.81 $\pm$ 0.04 (SN~1999aa)
to 1.93 $\pm$ 0.10 (SN~1991bg) for the \citet{Pri_etal06} templates.} 
these objects have been subjected to ever more intense
scrutiny.  \citet{Mei00}, \citet{Kri_etal04a}, \citet{Kri_etal04b, Kri_etal04c}, and
\citet{WV_etal08} have shown that over a wide range of decline rates the
near-IR absolute magnitudes of Type Ia SNe at some epoch with respect to
maximum light are essentially constant. In the near-IR most Type Ia SNe are
not just standardizable candles, but very nearly standard candles. The few
exceptions are mostly identifiable by their unusual IR light curve shapes.



Near-IR spectra of SN~2003gs \citep{Kot_etal09} and SN~1986G 
\citep{Fro_etal87} reveal the presence of lines from iron group elements that appeared earlier 
and  created stronger features than they do in spectra of normal Type Ia SNe.  $^{56}$Co is 
produced by the radioactive decay of $^{56}$Ni, which is the final burning product 
produced in the hottest and densest regions of the SN during the explosion.  The strong 
and early presence of iron group elements in SN 2003gs suggests that the outer layers of partially 
burned ejecta have lower opacities in SN~2003gs than they do in normal Type Ia SNe.  This 
is an important clue to understanding the photometric behavior of SN~2003gs, which 
indicates a faster than normal decline from approximately normal peak brightness.  In 
SN~2003gs, if the envelope surrounding the iron and cobalt regions is not as deep or more 
transparent than it is in normal Type Ia SNe, then $\gamma$ rays that power the 
luminosity will escape more easily and the decline rate will increase.


Here we present extensive optical ($UBVRI$) and near-IR ($YJHK$) data of the
nearby Type Ia SN~2003gs. The $Y$-band is a new photometric band
\citep{Hil_etal02} which exploits a relatively clean atmospheric window centred
at $\sim1.035\mu$m.

\section{Acquisition and reduction of photometry}
\label{reduction}

SN~2003gs was visually discovered by \citet {Eva03} on 2003 July 29.75 UT.
It was located at RA = 02:27:38.36, DEC = $-01\degr 09'35\farcs4$ (equinox 2000),
13\farcs4 east and 14\farcs6 south of the nucleus of the barred spiral
galaxy NGC~936.  A spectrum obtained with the CTIO 1.5-m telescope
on July 30.4 UT revealed it to be a subluminous Type Ia SN similar
to SNe~1991bg and 1999by at roughly 1 day before maximum light
\citep{Sun_etal03, MatSun_03}.  The spectroscopic typing was confirmed
using a spectrum obtained on July 31.33 with the 6.5-m Baade telescope
at Las Campanas Observatory \citep {Ham_etal03}. These authors also noted that
the spectrum lacked Na I D lines, suggesting that SN~2003gs is not
significantly reddened by dust.  Some preliminary
photometry was published in the IAU Circulars, including infrared
photometry accurate to $\pm$0.1 mag \citep{MikSzo_03}, but here we will only
consider our final, reduced photometry.
Figure \ref{finder} shows NGC 936, SN 2003gs, and the local sequence of stars.
A list of basic parameters is given in Table~\ref{properties}.

SN~2003gs is one of the few fast declining Type Ia supernova to be very well
observed both photometically and spectroscopically at optical and IR
wavelengths since SN~1986G \citep{Phi_etal87,Fro_etal87}.  Three
nights of infrared data of the prototypical fast decliner SN~1991bg were
published by \citet{Kri_etal04c}.  The fast decliner SN~1999by was
studied by \citet{Gar_etal04}; this object was observed on 14 nights
in the infrared, but the IR maxima were not well covered.

The fast decliners SNe 1991bg and 1999by are several tenths of a magnitude
fainter in the near-IR than more normal Type Ia SNe.  See Figure 16 of
\citet{Kri_etal04c}.  Given the rarity of fast declining Type Ia SNe,
SN~2003gs gives us an opportunity to study an object almost as extreme as
these two other examples.\footnote[16]{A slowly declining Type Ia SN has
decline rate \dmm\ $\lesssim$ 1.0.  A mid-range decliner has 1.0
$\lesssim$ \dmm\ $\lesssim 1.6$.  A fast decliner has \dmm\ $\gtrsim$ 1.6.
There are spectroscopic differences between the three groups.  The
slow decliners, for example, show the strongest lines due to doubly ionized
species and the weakest lines due to singly ionized species,
because they have hotter temperatures.}

Most of our photometry was obtained with the 1.3-m telescope at Cerro Tololo
Inter-American Observatory and the dual optical-infrared imager
ANDICAM.  The optical channel gives images with a scale of 0.369 arcsec
px$^{-1}$.  The gain is 2.3 electrons per analog-to-digital unit (ADU).
The readnoise is 6.5 electrons rms.  For the IR channel the plate
scale is 0.137 arcsec px$^{-1}$, the gain is 7.2 electrons per ADU, and
the readnoise is 20 electrons rms.  The optical photometry was derived using
point spread function (PSF) magnitudes.

The optical photometry was calibrated by first
determining the $UBVRI$ magnitudes of the field stars near SN~2003gs
and tying them to \citet{Lan92} standards.  One or two Landolt
fields were observed along with the SN~2003gs field on nine ostensibly
photometric nights using the CTIO 1.3-m.  The mean values of the photometry
of the secondary standards are given in Table \ref{optseq}.  Four nights
of optical photometry of the field stars, obtained with the 1-m Swope telescope
at Las Campanas Observatory (LCO) and the CTIO 0.9-m telescope, confirmed
that the $BVRI$ magnitudes of the field stars allowed photometric zeropoints to be
determined to better than $\pm$ 0.02 mag.  There may be systematic and random
errors in our $U$ band photometry at the 0.08 mag level owing to the
non-zero color terms on different systems and the inherently greater
scatter of $U$-band photometry.

Optical imagery of SN~2003gs was obtained with the 1-m Swope telescope
at LCO on 16 nights during the Carnegie Type II Supernova (CATS) Survey 
\citep{Ham_etal09}.  Reduction of the photometry from the LCO 1-m system 
had to include a term for each filter to account for non-linearities in
the response of the CCD camera \citep{Ham_etal06}.

A further five epochs of optical photometry were obtained with the
CTIO 0.9-m telescope from 2003 August 27 through 2004 February 16 UT.
The August 27 (CTIO) photometry was reduced
using PSF magnitudes.  For the final four epochs of CTIO 0.9-m imagery
and the final three epochs of CTIO 1.3-m imagery we subtracted template
images obtained with the CTIO 0.9-m on 2007 October 19 UT,
long after SN~2003gs had faded.

Some late time optical photometry was obtained with the University
of Arizona 1.54-m and 2.3-m telescopes.  The results were derived
using image subtraction templates obtained in November and December of 2005.

Experiments with the CTIO 0.9-m imagery with and without image subtraction
indicate no statistically significant differences through 57 days after
T($B_{max}$).  At $t$ = 91 days the $BVRI$ data obtained using image subtraction
are on average 0.03 mag fainter than PSF photometry without image subtraction.
These differences are comparable to the random errors of the photometry.

We observed two IR standards of \citet{Per_etal98}, P9104 and P9172, on 6
photometric nights along with the field of NGC~936 to calibrate the field
star immediately southeast of the SN (``star 3'' of the photometric
sequence). In Table \ref{field_star} we give the mean $YJHK$ values of this
field star, along with the $JHK$ values from the Two Micron All Sky Survey
(2MASS).  As one can see, the agreement is good.  Since not all the data were
taken on photometric nights when IR standards were observed, we derived
differential filter-by-filter magnitudes and added these differential values
to our derived photometry of the key field star in order to obtain the SN
photometry found in Table \ref{yjhk_data}.

We do not rely on measures of $Y$-band standards given
by \cite{Hil_etal02}.  Instead, we rely on synthetic photometry of
Sirius, Vega, and the Sun.  \citet{Kri_etal04b} give the following
relation:

\begin {displaymath}
(Y-K_s) \; = \; -0.013 \; + \; 1.614 \; (J_s-K_s) \; .
\end {displaymath}

This expression allows us to use the $J_s$ and $K_s$ magnitudes
of \citet{Per_etal98} standards to estimate the $Y$-band magnitudes
of those standards.  The $Y$-band calibrations should only be
considered approximate ($\pm$ 0.03 mag).

The $UBVRI$ and $YJHK$ photometry of SN~2003gs is given in
Tables \ref{ubvri_data} and \ref{yjhk_data}.  We note that the
$K$-band filter of ANDICAM on the CTIO 1.3-m is closer to the
$K$-short filter used at Las Campanas than a standard, wider
$K$-band filter.  Throughout this paper what we call $K$-band
photometry is really $K_s$ photometry.

Figure \ref{03gs_ubvri} shows the $UBVRI$ light curves.  For the
light curve fits we have corrected the ANDICAM $B$- and $V$-band photometry
to the filter system of \citet{Bes90}.  These are the so-called
S-corrections \citep{Str_etal02,Kri_etal03}.  Without these corrections the
$B-V$ colors of Type Ia SNe from ANDICAM are systematically
too red by as much as 0.1 mag.  

To determine the time of maximum light and the decline rate we used the
light curve analysis method of \citet{Pri_etal06}.  This method relies on
the well sampled light curves of 14 objects.  The two fastest decliners in
the training set are SNe 1992bo and 1991bg, which have \dmm\ = 1.69 and
1.93, respectively.  We find that SN~2003gs had a decline rate of \dmm\ =
1.83 $\pm$ 0.02.  The time of $B$-band maximum was JD 2452848.80 $\pm$ 0.53.
In Figure \ref{03gs_ubvri} we show the $BVRI$ templates corresponding to
\dmm\ = 1.83 and adjusted in the Y-direction to give the best fits.  Figure
\ref{resids} shows the residuals of the photometry with respect to the
family of \dmm\ = 1.83 templates. The $I$-band data, in particular, did not
correspond well to any template of \citet{Pri_etal06}.  We note with satisfaction,
however, that data from different telescopes are in reasonable agreement
with each other, better in fact than the agreement of the data
with the family of \dmm\ = 1.83 templates.

Figure \ref{03gs_yjhk} shows the $YJHK$ light curves.  The CTIO 1.3-m 
photometry includes the S-corrections given in Table \ref{irscorr}.
Given that we had access to IR spectra of SN~2003gs, it was possible to place
the CTIO 1.3-m IR photometry on the photometric system of \citet{Per_etal98} 
using the algorithm of \citet{Str_etal02} and \citet{Kri_etal03}.

According to the light curve fits, the first observations of SN~2003gs were
obtained {\em on} the date of the $V$-band maximum, some two days after
T($B_{max}$). We can say with certainty that the times of maximum light of
SN~2003gs in the $IJHK$ bands did {\em not} occur several days {\em after}
T($B_{max}$).  The timing of the IR maxima may be related to the range
of near-IR absolute magnitudes at maximum light (see \S \ref{photometry} 
below).

Figure \ref{03gs_bv} shows the $B-V$ colors of SN~2003gs, the zero reddening
line of \citet{Lir95}, and the $B-V$ color excess.  We note that Lira line
was derived from observations made with the CTIO 0.9-m.  We derive
statistically equivalent values of E($B-V$) using the two nights of CTIO
0.9-m photometry, or using them in combination with 8 nights of corrected
ANDICAM photometry, implying that the S-corrections were reasonably
appropriate.  Without adding the S-corrections to the ANDICAM $B$- and
$V$-band photometry, our value of E($B-V$) would be systematically too red by
about 0.1 mag \citep[see][Fig. 10]{Kri_etal03}.  The total $V$-band
extinction would then be systematically too large by $\sim$0.3 mag, given
standard dust \citep{Car_etal89}.

We find E($B-V$)$_{tot}$ = 0.066 mag, of which 0.035 mag is due to dust in
our Galaxy \citep{Sch_etal98}.  Thus, SN~2003gs is almost unreddened in its
host. Assuming R$_V$ = 3.1, we obtain total extinction corrections of 0.270,
0.205, 0.170, 0.124, 0.058, 0.037, and 0.024 mag for the $BVRIJHK$ bands,
respectively, using the scale factors given by \citet[][Table
8]{Kri_etal06}.  As we have found before, the uncertainties in the near-IR
extinction corrections are comparable to the random errors of the
photometry, even if a given object is dimmed and reddened by dust with
non-standard properties.

In Table \ref{properties} we give the absolute magnitudes of SN~2003gs for
the $BVRI$ and $JHK$ bands.  These rely on the distance modulus of NGC~936
of $m-M$ = 31.81 $\pm$ 0.28 mag given by \citet{Ton_etal01}, based on the
method of surface brightness fluctuations (SBFs), and corrected by 0.16 mag
to $m-M$ = 31.65 to account for a systematic error in the $I$-band Cepheid
period-luminosity relation used for the calibration of the SBF distances
\citep{Jen_etal03}. We have adopted the total extinction values given above.

\citet[][Eqns. 2, 3, and 4]{Gar_etal04} give exponential fits to the $BVI$
decline rate relations, which fit the absolute magnitudes over the
full range of decline rates of Type Ia SNe.  While SN~2003gs was subluminous
compared to mid-range decliners at optical wavelengths, we find
that SN~2003gs was 0.18, 0.15, and 0.30 mag brighter in the $B$-, $V$-, and
$I$-bands, respectively, than the values implied by the relationships
of \citet{Gar_etal04} for a Type Ia SN with \dmm\ = 1.83.

The $J$-band data just after maximum light are convincingly
fit with the {\em unstretched} template given by \citet{Kri_etal04b}.
The implication is that the $J$-band maximum was 0.23 mag brighter than our
first (S-corrected) value, obtained 2.1 days after T($B_{max}$).  That the
unstretched template fits the data is evidence that in the $J$-band
the photometric behavior of SN~2003gs was more like a mid-range decliner
than a fast decliner.  If we use the unstretched maximum light
$H$- and $K$-band templates of \citet{Kri_etal04b}, the $H$- and $K$-band
maxima were approximately 0.13 mag brighter than our earliest observations.
Along with the small near-IR extinctions and a distance modulus of 31.65 mag,
the resulting absolute magnitudes are M$_J$ = $-$18.50, M$_H$ = $-$18.48, and
M$_K$ = $-$18.37.  These values are 0.11 mag fainter in $M_J$, 0.18 mag
brighter in $M_H$, and 0.07 mag fainter in $M_K$ than the mean values
of the slow decliners and mid-range decliners (see below).

In Figure \ref{vjhk} we show the unreddened $V$ minus near-IR colors of
SN~2003gs and several other Type Ia SNe: the very normal mid-range decliner
2001el \citep{Kri_etal03}; and the fast decliners 1999by \citep{Gar_etal04},
2005bl \citep{Tau_etal08, WV_etal08}, 2005ke, and 2006mr \citep{Con_etal09,
Fol_etal09}.  We have also used the $K$-band data of SN~2005ke from
\citet{WV_etal08}.  SNe 199by, 2003gs, 2005bl, and 2005ke show similar $V-H$
and $V-K$ colors from the earliest times to $t \sim$ +16 days. There is a
ridge line onto which $V-H$ and $V-K$ colors converge, starting at $t \sim
+18.5$ d for the fast decliners; these loci meet up with the SN~2001el data
starting at $t \sim$ 30 d.

In Table \ref{vmircolors} we give low order polynomial fits to the early-
and late-time $V-H$ and $V-K$ colors shown in Fig.  \ref{vjhk}. The
early-time colors of SNe 2001el and 2006mr have been excluded from the fits.  
Given the reduced $\chi ^2$ values and the $\pm$0.2 mag rms scatter of the 
fits to the early-time data of SNe
1999by, 2003gs, 2005bl, and 2005ke, we should not consider these colors to
be ``uniform''.  But we find that the $V-H$ and $V-K$ colors of the fast
decliners are remarkably uniform after $t \sim$ +30 d, and identical to the
unreddened colors of the mid-range decliner SN~2001el.  The rms scatter in
the two color indices for the linear decline is $\pm$0.1 mag. We are
reminded of the so-called ``Lira law'' for the $B-V$ colors of Type Ia SNe
starting at $t \sim$ +32 d.  The only late-time outliers in the $V-H$ and
$V-K$ plots are the $V-H$ data of SN~2003gs, which are apparently redder than
other objects due to an interesting bump in the $H$-band spectra of SN~2003gs
after $t \sim$ +45 d \citep{Kot_etal09}.

\section{Discussion}
\subsection{Photometry}\label{photometry}
\subsubsection{Light curve morphology}

\citet{Ham_etal96} first showed that a stack of
$I$-band light curves, ordered by the optical decline rate parameter,
exhibits weaker and weaker secondary maxima as we proceed from the slowest
to the fastest decliners. If we consider the mean flux 20 to 40 days after
the time of $B$-band maximum, typical mid-range decliners have a secondary
maximum that is 0.5 times as strong as the I-band maximum \citep[][Fig.
17]{Kri_etal01}. For the fast decliners the second flux peak is weak enough
that it just blends in with the principal decline.

SN~2003gs exhibited no secondary hump in the $I$-, $H$- and $K$-bands. There
is only a weak $J$-band secondary hump.  In the $Y$-band, however, SN~2003gs
had a secondary maximum that was brighter than the SN must have been at the
time of $B$-band maximum.  This was also the case with the mid-range decliner
SN~2000bh \citep{Kri_etal04b}. Many Type Ia SNe observed by the Carnegie Supernova
Project\footnote[17]{http://csp1.lco.cl/~cspuser1/PUB/CSP.html} (CSP) show the
same phenomenon \citet{Con_etal09}. The secondary hump is apparently
maximized in the 1.03 $\mu$m band.  According to \citet{Kas06}, it is due to
a change of opacity, when the expanding fireball undergoes a transition from
primarily doubly ionized species to singly ionized ones.

\citet{Mil_etal01} studied the late time light curve evolution of Type Ia
SNe. From 50 to 200 days after the time of explosion (i.e. about 30 to 180
days after the time of $B$-band maximum) normal Type Ia SNe exhibit linearly
declining light curves, with rates of decline of 1.43, 2.12, 2.62, and 2.67
mag per 100 days in the $BVRI$ bands, respectively.  In Fig. \ref{late_time}
we show the differences of the photometry with respect to these nominal rates
of decline. The 50 to 200 day rates of decline for SN~2003gs are similar, but
not exact matches, to the other six submulinous objects.  The $B$- and
$V$-band rates of decline for the fast declining objects differ appreciably
from the normal events, but for the $R$- and $I$-bands there is less
variation.

Two excellent reviews of the photometric and spectroscopic properties of
fast declining Type Ia SNe have been published by \citet{Gar_etal04} and
\citet{Tau_etal08}.   Of greatest interest to us here is the intrinsic
brightness of SN~2003gs and other fast decliners.

\subsubsection{Fast decliners to consider}
\label{fast_decliners}

\citet{Kri_etal04a} and \citet{Kri_etal04c} have already considered the
absolute peak magnitudes in the near-IR of SNe 1986G \citep{Fro_etal87,
Phi_etal87}, 1991bg \citep {filippenko:92b, leibundgut:93} and 1999by
\citep{Gar_etal04}. Here we use a corrected distance modulus to NGC 5128 (the
host of SN~1986G) of $m-M = 27.90 \pm 0.14$ mag by subtracting 0.16 mag
\citep{Jen_etal03} from the SBF distance modulus of \citet[][Table
3]{Ajh_etal01}.  This is related to a systematic error in the calibration of
the $I$-band Cepheid period-luminosity relation used to anchor the SBF
distances.  Like SN~2003gs, SN~1986G appears to be a fast decliner that did
not peak ``late'' in the near-IR. One concern is that this object suffered
significant extinction.  However, from polarimetry data we know that
that R$_V$ = 2.4 is the appropriate value to used for dust in the host
of SN~1986G \citep{Hou_etal87}.  As with many objects, a larger source of
uncertainty in the near-IR absolute magnitudes comes from the uncertainty
of the distance modulus, not the uncertainty in the near-IR extinction
corrections.

Other fast decliners we can consider are these:

{\em SN~2003hv}. \citet{Lel_etal09} present spectra and extensive photometry.  
It had a decline rate of \dmm\ = 1.61 $\pm$ 0.02.  The distance modulus is
$m-M$ = 31.53 $\pm$ 0.30 mag \citep{Ton_etal01}, which becomes $m-M$ = 31.37
mag after applying the correction of \citet{Jen_etal03}.  This object was
unreddened in its host.  We estimate that SN~2003hv peaked in the $J$-band 
1.7 days prior to T($B_{max}$).  The absolute magnitudes at maximum were
M$_J \approx -18.52$, M$_H \approx -18.17$, and M$_K \approx -18.33$, 
with uncertainties of $\pm$ 0.31 mag.  The peak brightness
and the application of Arnett's Law \citep{arnett:82} suggest that it produced 
0.40 to 0.42 M$_{\odot}$ of $^{56}$Ni, somewhat more than was produced by 
the fastest decliners (see Table \ref{nickel}).

{\em SN~2004gs} \citep{Con_etal09, Fol_etal09}.  \dmm\ = 1.54 $\pm$ 0.01.
This was a reasonably fast decliner, but the first IR data were obtained
at +6.6 d after T($B_{max}$).  We do not know if it peaked early or late,
and to extrapolate back to the IR maxima assumes that its light curve
obeyed templates based on other objects.  

{\em SN~2005bl} \citep{Tau_etal08, Fol_etal09, WV_etal08}.  We adopt \dmm\ =
1.80 $\pm$ 0.04 \citep{Fol_etal09}.  \citet{Tau_etal08} give \dmm\ = 1.93
$\pm$ 0.10.  The $IJHK$ maxima occurred a few days after T($B_{max}$).
The distance modulus is $m-M$ = 35.10 $\pm$ 0.09 mag, given the
radial velocity in the frame of the Cosmic Microwave Background, v$_{CMB}$ =
7534 \kms Mpc$^{-1}$, and a Hubble constant of 72 \kms\ Mpc$^{-1}$
\citep{Fre_etal01}.  E($B-V$)$_{tot} \approx$ 0.20 $\pm$ 0.08, giving $JHK$
extinctions of 0.17, 0.11, and 0.07 mag, respectively, with uncertainties
of 40 percent.  Using the $JHK$ apparent magnitudes at maximum of
\citet{WV_etal08}, these extinctions, and the Hubble flow distance
modulus, we obtain absolute magnitudes at maximum light of
M$_J$ = $-17.92 \pm 0.12$,
M$_H$ = $-17.88 \pm 0.12$, and
M$_K$ = $-17.70 \pm 0.17$.

{\em SN~2005ke} \citep{Con_etal09, Fol_etal09}. \dmm\ = 1.76 $\pm$ 0.01. The
near-IR maxima occurred 1-2 days after T($B_{max}$). Its host was NGC 1371,
a member of the Eridanus group. \citet[][Table 4]{Ton_etal01} give an SBF
distance modulus of $m-M$ = 32.00 $\pm$ 0.08 mag for the group, which
becomes $m-M$ = 31.84 mag after the \citet{Jen_etal03} correction.  Using the
zero reddening line of \citet{Lir95}, we find E($B-V$)$_{tot}$ = 0.066,
giving $JHK$ extinctions of 0.058, 0.037, 0.024 mag.  The Las Campanas
photometry gives $J_{max}$ = 14.00 $\pm$ 0.02, $H_{max}$ = 13.95
$\pm$ 0.03.  We adopt $K_{max}$ = 14.03 $\pm$ 0.02 \citep{WV_etal08}.  The
resulting absolute magnitudes are M$_J$ = $-$17.90, M$_H$ = $-$17.93, and
M$_K$ = $-$17.83, to which we assign conservative uncertainties of $\pm$
0.24 mag.  We note that SN~2005ke showed evidence of interacting with the
nearby circumstellar medium, based on X-ray observations with {\em Swift} \citep{Imm_etal06}.

{\em SN~2006gt} \citep{Con_etal09, Fol_etal09}.  \dmm\ = 1.66 $\pm$ 0.03.
This was a distant fast decliner (redshift 0.0448).
The $J$-band maximum clearly occurred prior to T($B_{max}$). 
If we simply take the earliest available IR observations
and apply no extinction corrections at all, we get M$_J$ = $-$18.45 and
M$_H$ = $-$18.21.  This appears to be another object that peaked early
and was not faint.

{\em SN~2006mr} \citep{Con_etal09, Fol_etal09}.  \dmm\ = 1.82 $\pm$ 0.02.
The near-IR maxima occurred 3-4 days after T($B_{max}$).  Its host was
NGC~1316 (Fornax A), whose distance modulus is $m-M$ = 31.59 $\pm$ 0.08 mag
\citep{Can_etal07}.  Using the zero reddening line of \citet{Lir95} implies
that this object has negative reddening, so we shall adopt E($B-V$)$_{Gal}$
= 0.021 \citep{Sch_etal98} as the color excess.  The implied extinctions in
the $J$- and $H$-bands are then 0.018 and 0.012 mag, respectively. The Las
Campanas photometry gives $J_{max}$ = 13.99 $\pm$ 0.03, $H_{max}$ = 13.85
$\pm$ 0.04.  The resulting absolute magnitudes are M$_J$ = $-$17.62 $\pm$
0.10 and M$_H$ = $-$17.75 $\pm$ 0.10.

We do not consider the fast decliner SN~2000bk, which had a decline rate of
\dmm\ = 1.63 \citep{Kri_etal01}. The first optical photometry was only
obtained 10.8 days after the derived time of $B$-band maximum.  The first
near-IR data were obtained at $t$ = +5.9 d.  We do not know if it peaked
early or late in the near-IR compared to $B$.  Also, the $H$-band data were
somewhat ragged.

Also, we do not consider the unusual SN~2005hk \citep{Phi_etal07}.  Along
with SN~2002cx \citep{Li_etal03}, it may belong to a new subclass of Type Ia
SNe. In their paper on the very subluminous SN~2008ha \citet[][]{RJF_etal09}
list 14 objects similar to SN~2002cx.  \citet{Val_etal09} even suggest that
these SNe are core collapse objects, not Type Ia SNe.

In Fig. \ref{jband} we show the $J$-band photometry of six fast declining
Type Ia SNe out to $t$ = 35 d.  The top four objects are consistent with
the stretchable template of \citet{Kri_etal04b}.  They peak prior
to T($B_{max}$).  SNe~2005ke and 2006mr, on the other hand, cannot be fit
with same template no matter how it is stretched.  That is because
these objects peaked in the $J$-band after T($B_{max}$).  Note also that 
the late peakers have much weaker secondary maxima.  These are two 
of five known fast decliners that have fainter IR absolute magnitudes at maximum.

\subsection{A bimodal distribution of absolute magnitudes at maximum}
\label{bimodal}

In Fig. \ref{absmags} we show the near-IR absolute magnitudes
from Fig. 16 of \citet{Kri_etal04c}, along with
SN~2000cf \citep{Kri_etal06}, SN~2004S \citep{Kri_etal07},
SN~2003gs, and the fast decliners mentioned above.  We have made
the values of SN~1980N brighter by 0.15 mag, as we previously had adopted a
distance modulus of 31.44 mag for its host, which happens to be the
same host as that of SN~2006mr.

At first glance Fig. \ref{absmags} indicates that there is just some range of
near-IR absolute magnitudes of Type Ia SNe at maximum light.  But taking the
data at face value, we know of four fast declining Type Ia SNe (1986G,
2003gs, 2003hv, and 2006gt)  that peaked early in the near-IR {\em and} whose near-IR
absolute magnitudes at maximum light were {\em not faint}.  We may also 
consider five fast declining Type Ia
SNe that peaked late in the near-IR and whose near-IR absolute magnitudes at
maximum light were on average significantly fainter than the mean values for
Type Ia SNe that peak $\sim$2-3 days before T($B_{max}$). Those late peaking
fast decliners are, on average, 0.81 mag fainter in $M_J$, 0.44 mag fainter
in $M_H$, and 0.64 mag fainter in $M_K$ than the early peakers.  We suggest
that there is a bimodal distribution of maximum-light near-IR absolute
magnitudes of Type Ia SNe.  Which group a particular object falls into
depends on whether it peaks early or late compared to T($B_{max}$).  A
summary of the absolute magnitudes of the two groups is given in Table
\ref{absmag_avgs}.

In Fig. \ref{early_late} we present histograms of the absolute magnitudes
shown in Fig. \ref{absmags}.  For the $J$- and $K$-band data there
is a noticeable bimodal distribution, depending on whether the objects
peak early or late with respect to T($B_{max}$).  In the $H$-band the
two groups are not disconnected.

Another way of presenting the data for the fast decliners is shown in
Figure \ref{delay}, where we plot the $J$-band absolute magnitudes at
maximum light vs. the difference in the times of maximum light in
the $J$-band and the $B$-band.  A regression line is fit to seven
of the nine points.  For SN~1991bg, on the basis of scanty data,
T($J_{max}$) occurred 2.1 days or less after T($B_{max}$).  For
SN~1999by we assume that the $J$-band maximum occurred at the same time
as the $I$-band maximum, which was well covered \citep{Gar_etal04}.

An easy way to understand the ``early and bright'' vs. ``late and faint'' 
situation is as follows.  For a (brighter) SN which has two distinct
humps in the light curve these two humps occur at times $t_1$ and $t_2$.
As long as there are two humps, $t_{max} = t_1$.
As we proceed to weaker explosions ($t_2 - t_1$) decreases. 
Finally, when $t_2$ occurs soon after $t_1$ the two humps
merge.  In this case what we observe is a single hump whose
maximum occurs at

\begin {displaymath}
t^{\prime}_{max} \; = \frac{\alpha t_1 \; + \; t_2}{1 \; + \; \alpha}  \; ,
\end {displaymath}

\parindent = 0 mm

where $\alpha \gtrsim 2$ for the $I$- and $J$-band light curves. 
Details on modelling the secondary maxima are discussed by \citet[][\S3]{Hoe_etal95}
and \citet{Kas06}.

\parindent = 9 mm

{\em All} of the slow decliners and mid-range decliners that have been
observed early peaked early.  Only the fast decliners that peak late appear
to be fainter in the near-IR.  We make two predictions: 1) Type Ia SNe that peak 
early are standard candles in the near-IR at peak brightness; and 2) Type Ia SNe that
peak a few to several days after T($B_{max}$) in the near-IR are subluminous
in all bands. We would, of course, be interested in the photometric and
spectroscopic properties of any slow decliner or mid-range decliner that
peaks several days after T($B_{max}$).  So far no such object has been
observed.

On the basis of the data for nine fast declining Type Ia SNe we assert that
the absolute magnitudes at maximum light in the near-IR are related 
to the relative times of the near-IR maxima and T($B_{max}$).  
A considerably larger dataset is required to demonstrate for certain if it
is a bimodal distribution or a linear trend.  

\subsection{Bolometric light curve and the mass of $^{56}$Ni}
\label{bolo_lc}

Using the $UBVRIJHK$ fluxes, the value of the distance modulus
listed in Table \ref{properties}, and extinction given in
\S \ref{reduction}, we constructed
an optical-near-infrared (OIR) pseudo-bolometric light curve,
shown in Fig. \ref{blc_jensen}. We included a correction for the
UV flux as described by \citet{suntzeff:96}.

The peak of the bolometric light curve is directly related to the amount of
$^{56}$Ni produced in the explosion. It is well known that the peak
quasi-bolometric luminosities of Type Ia SNe span a range of at least a factor
of 25, implying a fairly large range of $^{56}$Ni masses. Although our observations
of SN~2003gs began after maximum light, we can still estimate the amount of
$^{56}$Ni, bearing this caveat in mind. To do so, we employ Arnett's rule
\citep{arnett:82} as parametrized by \citet[][their equation 7]{strtz:05}, and
obtain a $^{56}$Ni mass of about 0.25$\,M_\odot$. \citet{strtz:05} assumed a
bolometric rise-time for normal Type Ia SNe (\dmm $\sim$ 1.1) to be 19 days.
This value may not be quite appropriate for objects like SN~2003gs, which have
fast-evolving light curves. 

Recently, \citet[][their Fig. 6]{Tau_etal08} collected the available photometry
for a number of Type Ia SNe having $1.88 \lesssim \dmm \lesssim 1.95$ and
compiled quasi-bolometric light curves for these. For the fast decliners in
their sample, the peak luminosities ranged from $\sim 10^{42.05} - 10^{42.3}$
erg\,s$^{-1}$. We report the resulting values of the $^{56}$Ni mass as obtained
using the parametrization described above in Table \ref{nickel}, and find
that all of them cluster in the 0.1\,$M_\odot$ range. 

Interestingly, the narrow and rapidly declining light curve of SN~2003gs as
indicated by its \dmm\ of 1.83, belies the lower limit to its peak luminosity
($\sim 10^{42.7}$ erg\,s$^{-1}$). This may be taken to be further evidence that
Type Ia SNe do not form a one parameter family.

\subsection{Spectroscopy}

Optical spectra of SN~2003gs \citep{Sun_etal03, MatSun_03, Ham_etal03}
obtained with the CTIO 1.5-m telescope and also the Magellan telescopes soon after 
T($B_{max}$) show the presence of Ti~II in the region $4000-4500 \AA$ and and a large 
ratio of Si~II at 5800$\AA$ to Si~II at 6150$\AA$ that are characteristics of moderately 
fast declining Type Ia SNe as described in \citet{Gar_etal04}.  

These spectra are shown in Figure \ref{spectra}.  Using the Supernova Identification
code SNID \citep{BloTon07}, we find that the two spectra obtained near maximum light
are very similar to those of SN~2004eo \citep{pastorello:07b} at a comparable epoch.
SN~2004eo had \dmm\ = 1.46 and produced 0.45 M$_{\odot}$ of $^{56}$Ni, similar to SN~2003hv.
At $t$ = 18 and 49 days, however, our optical spectra of SN~2003gs are similar to
the prototypical fast decliner SN~1991bg.  Many more optical spectra of SN~2003gs 
are presented by \citet{Kot_etal09}.

Near-infrared (NIR; 0.8-2.5 $\mu$m) spectra provide a rich source of information about 
the physical characteristics of SNe Ia because many elements are undetectable or obscured 
by line blending at other wavelengths but produce strong lines in the NIR 
\citep{Mei_etal96,wheeler:98,bowers:97,pah:02,marion:03, marion:09}.  NIR spectroscopic 
observations are particularly effective for characterizing the chemical structure of the 
supernova at different layers because the optical depth for most lines is smaller in the 
NIR than at shorter wavelengths so that a greater radial depth can be probed with each 
spectrum \citep{wheeler:98}.

\citet{Fro_etal87} present a complete near-IR spectrum of SN~1986G obtained at 
+12 d and three partial spectra through +24 d.  The data of SN~1986G closely 
resemble the spectra of SN~2003gs obtained at similar epochs.  \citet{Kot_etal09} 
present 11 near-IR spectra of SN 2003gs obtained between +4 d to 91 d after 
T($B_{max}$).  These authors kindly loaned us their spectra for the calculation
of near-IR S-corrections to the photometry.  We will not give an analysis
of their spectra here, except to note that strong lines of iron group elements
are seen early in the sequence.

The strong and early presence of iron and cobalt in the spectra of SN 2003gs 
indicates a lower opacity in the covering layers for SN~2003gs than is found in 
normal Type Ia SNe.  This could be due to asymmetries in the explosion that 
place the iron and cobalt line-forming regions physically closer to the surface of the SN 
or to an explosion that produced a larger quantity of $^{56}$Ni.  It is also 
possible that the depth of the surrounding envelope is approximately the same, 
but the opacities at these wavelengths are reduced by some other mechanism.  In 
\S \ref{bolo_lc} we showed that SN~2003gs has a larger $^{56}$Ni mass than 
other fast declining Type Ia SNe (see Table \ref{nickel}).  In any case, this 
is an important clue to understanding the photometric behavior of SN~2003gs, 
which indicates a faster than normal decline from approximately normal peak 
brightness.  If the envelope surrounding the iron and cobalt line-forming regions 
is not as deep or is more transparent than in normal Type Ia SNe, then $\gamma$ rays 
that power the luminosity will escape more easily and the decline rate will 
increase.


\section{Conclusions}

We have presented one of the most complete photometric datasets available for a 
fast declining Type Ia SN.  SN~2003gs was first observed on the date of 
$V$-band maximum light, some two days after $B$-band maximum.  Our coverage 
continued without any serious gaps until $t$ = 91 d.  We also obtained some 
late time photometry.

We deduce that SNe 1986G, 2003gs, 2003hv, and 2006gt were fast declining 
objects that shared some interesting photometric characteristics.  These 
objects were subluminous in the optical band passes, but their near-IR maximum 
light absolute magnitudes were statistically equal to those of the slow 
decliners and mid-range decliners. Also, the near-IR maxima of these four
objects apparently occurred prior to the time of $B$-band maximum light.  In 
the case of SN~2003gs we can say that the near-IR maxima did {\em not} occur 
``late,'' i.e. a few days after T($B_{max})$. Type Ia SNe that had late near-IR 
maxima (SNe 1991bg, 1999by, 2005bl, 2005ke, and 2006mr) were subluminous at the 
times of the IR maxima.  There appears to be a bimodal distribution of 
near-IR absolute magnitudes of Type Ia SNe at maximum light.  Which group a 
particular object falls into depends on whether it peaked late or early.  This 
empirical finding is undoubtedly related to the opacity in the expanding 
fireball, and should help us refine models of Type Ia SNe.

Near-IR spectral data appear to show that NIR opacities in the outer layers are 
lower in SNe 1986G and 2003gs than they are in normal Type Ia SNe.  If the 
$\gamma$ rays are less confined, then the observed luminosity decline rate 
would be accelerated.  That result is consistent with the photometric result 
for these objects that indicates a faster than normal decline from 
approximately normal peak brightness in the NIR.


\acknowledgments

We thank Rubina Kotak for access to IR spectra of SN~2003gs ahead of publication
and for calculating the bolometric light curve of SN~2003gs.
E. A. K. was supported by the REU program of the National Science
Foundation. We particularly thank the Carnegie Supernova Project for access to data
prior to publication.  We thank Jose Luis Prieto for his $BVRI$ light curve
fitting templates.  We thank Peter Hoeflich for useful discussions relating to
the ``late and faint'' effect.  We made use of the
NASA/IPAC Extragalactic Database (NED), data of the Two Micron All Sky
Survey, and SIMBAD, operated at CDS, Strasbourg, France.  Most of the optical
and IR photometry was obtained with the CTIO 1.3-m telescope, which is
operated by the SMARTS consortium.  Without the rapid response made
possible by SMARTS, this valuable dataset would not have been obtained.
M.H. and R.C. acknowledge support provided by FONDECYT through grant
1060808, the Millennium Center for Supernova Science through grant
P06-045-F, Centro de Astrof\'{i}sica FONDAP 15010003, Center of
Excellence in Astrophysics and Associated Technologies (PFB 06).
R. C. was supported by CONICYT through Programa Nacional de Becas de
Postgrado grant D-2108082.


\clearpage

\begin{deluxetable}{ll}
\tablewidth{0pt}
\tablecolumns{2}
\tablecaption{Properties of SN~2003gs and its host galaxy\label{properties}}
\tablehead{
\colhead{Parameter} & \colhead{Value} 
}
\startdata
Host galaxy                                       & NGC 936 \\
Host galaxy type\tablenotemark{a}                 & SB0/SBa      \\
Heliocentric radial velocity\tablenotemark{a}     & $1430$\,\kms  \\
Distance modulus\tablenotemark{b}                 & $31.65\pm 0.28$ mag          \\
E$(B-V)_{Gal}$\tablenotemark{c}                   & 0.035 $\pm$ 0.003 mag \\
RA of SN (J2000)                                  & $2^h 27^m 38^s\hspace{-1 mm}.36$  \\
Dec of SN (J2000)                                 & $-01\degr 09'35\farcs4$  \\
Offset from nucleus                               &  13\farcs4\,E \, 14\farcs6\,S  \\
Julian Date of $B$-band maximum                   &  $2452848.80 \pm 0.53$ \\
$\Delta m_{15}$(B)$$                              & 1.83 $\pm $0.02    \\
M$_{B,max}$                                       & $-17.94 \pm 0.29$ \\
M$_{V,max}$                                       & $-18.38 \pm 0.29$ \\
M$_{R,max}$                                       & $-18.53 \pm 0.29$ \\
M$_{I,max}$                                       & $-18.45 \pm 0.29$ \\
M$_{J,max}$                                       & $-18.50 \pm 0.29$ \\
M$_{H,max}$                                       & $-18.48 \pm 0.29$ \\
M$_{K,max}$                                       & $-18.37 \pm 0.29$ \\
\enddata
\tablenotetext{a}{From NED.}
\tablenotetext{b}{\citet{Ton_etal01}, using method of Surface Brightness 
Fluctuations, and corrected by 0.16 mag \citep{Jen_etal03}.}
\tablenotetext{c}{\citet{Sch_etal98}.}
\end{deluxetable}

\begin{deluxetable}{rccccc}
\tablewidth{0pt}
\tablecolumns{6}
\tablecaption{Optical Field Star Sequence near SN~2003gs\label{optseq}}
\tablehead{
\colhead{ID}  &
\colhead{$U$} &
\colhead{$B$} &
\colhead{$V$} &
\colhead{$R$} &
\colhead{$I$}
}
\startdata

 1 & 16.332  & 15.450 & 14.425 & 13.775  & 13.211 \\
 2 & 14.072  & 13.957 & 13.314 & 12.945  & 12.557 \\
 3 &  \ldots & 17.890 & 16.271 & 15.235  & 13.976 \\
 4 &  \ldots & 18.125 & 16.575 & 15.520  & 14.147 \\
 5 & 18.444  & 18.334 & 17.639 & 17.210  & 16.762 \\
20 & 18.016  & 17.160 & 16.195 & 15.596  & 15.060 \\
\enddata
\end{deluxetable}

\begin{deluxetable}{ccc}
\tablewidth{0pt}
\tablecolumns{3}
\tablecaption{Near-Infrared Photometry of Field Star Near SN~2003gs\tablenotemark{a}\label{field_star}}
\tablehead{
\colhead{Band} &
\colhead{Derived\tablenotemark{b}} &
\colhead{2MASS values}
}
\startdata
$Y$ & 13.221 (0.016) &   \ldots       \\
$J$ & 12.779 (0.011) & 12.812 (0.027) \\
$H$ & 12.258 (0.009) & 12.251 (0.022) \\
$K$ & 11.985 (0.014) & 11.993 (0.019) \\
\enddata
\tablenotetext{a}{RA = 02:27:40.63, DEC = $-$01:10:05.1 (equinox 2000).  This is
``star 3'' of the field star sequence.}
\tablenotetext{b}{With respect to stars 9104 and 9172 of \citet{Per_etal98}.}
\end{deluxetable}

\begin{deluxetable}{lccccccc}
\tablewidth{0pt}
\tablecolumns{8}
\tabletypesize{\scriptsize}
\tablecaption{Optical Photometry of SN~2003gs\label{ubvri_data}}
\tablehead{
\colhead{JD\tablenotemark{a}} &
\colhead{Epoch\tablenotemark{b}} &
\colhead{$U$} &
\colhead{$B$} &
\colhead{$V$} &
\colhead{$R$} &
\colhead{$I$} &
\colhead{Telescope\tablenotemark{b}}
}
\startdata
2850.80   & \phn\phn+2.0    &  \ldots          & 14.125 (0.016) & 13.491 (0.014) & 13.249 (0.015) & 13.253 (0.014) & 2 \\ 
2850.91   & \phn\phn+2.1    &   14.731 (0.041) & 14.134 (0.012) & 13.488 (0.009) & 13.279 (0.008) & 13.387 (0.013) & 1 \\
2851.80   & \phn\phn+3.0    &  \ldots          & 14.235 (0.016) & 13.511 (0.014) & 13.263 (0.015) & 13.244 (0.014) & 2 \\
2852.87   & \phn\phn+4.1    &   14.993 (0.012) & 14.356 (0.007) & 13.505 (0.006) & 13.311 (0.006) & 13.399 (0.009) & 1 \\
2854.88   & \phn\phn+6.1    &   15.303 (0.021) & 14.644 (0.013) & 13.614 (0.011) & 13.414 (0.003) & 13.462 (0.010) & 1 \\
2854.88   & \phn\phn+6.1    &     \ldots       & 14.651 (0.018) & 13.604 (0.015) & 13.418 (0.012) & 13.467 (0.014) & 1 \\
2856.86   & \phn\phn+8.1    &   15.761 (0.083) & 14.982 (0.018) & 13.781 (0.012) & 13.544 (0.011) & 13.503 (0.014) & 1 \\
2858.86   & \phn+10.1       &   16.031 (0.015) & 15.333 (0.009) & 13.955 (0.006) & 13.663 (0.006) & 13.519 (0.005) & 1 \\
2861.88   & \phn+13.1       &   16.428 (0.015) & 15.762 (0.009) & 14.226 (0.006) & 13.839 (0.006) & 13.565 (0.010) & 1 \\
2864.86   & \phn+17.1       &   16.587 (0.042) & 16.047 (0.019) & 14.515 (0.011) & 14.080 (0.011) & 13.699 (0.012) & 1 \\
2866.90   & \phn+18.1       &  \ldots          &  \ldots   	& 14.730 (0.015) &  \ldots        &    \ldots      & 6 \\
2867.80   & \phn+19.0       &   16.890 (0.050) & 16.297 (0.012) & 14.826 (0.006) & 14.396 (0.005) & 13.993 (0.005) & 1 \\
2870.80   & \phn+22.0       &  \ldots          & 16.489 (0.015) & 15.178 (0.014) & 14.675 (0.014) & 14.235 (0.014) & 2 \\
2872.74   & \phn+23.9       &   17.258 (0.144) & 16.607 (0.020) & 15.202 (0.013) & 14.833 (0.012) & 14.438 (0.015) & 1 \\
2876.80   & \phn+28.0       &   17.274 (0.013) & 16.718 (0.013) & 15.394 (0.013) & 15.061 (0.013) & 14.715 (0.014) & 1 \\
2878.75   & \phn+30.0       &   17.136 (0.025) & 16.724 (0.024) & 15.475 (0.015) & 15.100 (0.011) & 14.702 (0.019) & 3 \\
2882.75   & \phn+34.0       &   17.434 (0.051) & 16.870 (0.016) & 15.615 (0.011) & 15.345 (0.011) & 15.014 (0.014) & 1 \\
2885.73   & \phn+36.9       &   17.505 (0.057) & 16.912 (0.021) & 15.730 (0.011) & 15.469 (0.010) & 15.138 (0.014) & 1 \\
2888.90   & \phn+40.1       &  \ldots          & 16.957 (0.015) & 15.904 (0.014) & 15.575 (0.014) & 15.252 (0.014) & 2 \\
2890.78   & \phn+42.0       &   17.769 (0.062) & 17.069 (0.022) & 15.887 (0.011) & 15.677 (0.010) & 15.387 (0.015) & 1 \\
2890.80   & \phn+42.0       &  \ldots          & 17.018 (0.015) & 15.941 (0.014) & 15.635 (0.014) & 15.334 (0.014) & 2 \\
2897.78   & \phn+49.0       &   18.062 (0.088) & 17.145 (0.021) & 16.093 (0.016) & 15.937 (0.014) & 15.657 (0.015) & 1 \\
2904.73   & \phn+55.9       &   18.345 (0.145) & 17.281 (0.041) & 16.333 (0.017) & 16.192 (0.014) & 15.958 (0.028) & 1 \\
2905.70   & \phn+56.9       &  \ldots          & 17.244 (0.016) & 16.376 (0.015) & 16.181 (0.015) & 15.957 (0.014) & 2 \\
2905.76   & \phn+57.0       &   17.952 (0.090) & 17.224 (0.036) & 16.381 (0.040) & 16.260 (0.033) & 16.079 (0.033) & 3 \\
2906.80   & \phn+58.0       &  \ldots          & 17.264 (0.016) & 16.451 (0.015) & 16.260 (0.014) & 15.998 (0.014) & 2 \\
2907.80   & \phn+59.0       &  \ldots          & 17.277 (0.019) & 16.473 (0.022) & 16.323 (0.025) & 16.056 (0.020) & 2 \\
2908.80   & \phn+60.0       &  \ldots          & 17.298 (0.015) & 16.459 (0.014) & 16.312 (0.014) & 16.061 (0.015) & 2 \\
2911.75   & \phn+63.0       &   18.400 (0.039) & 17.374 (0.015) & 16.560 (0.011) & 16.487 (0.011) & 16.215 (0.012) & 1 \\
2914.80   & \phn+66.0       &  \ldots          & 17.395 (0.019)	& 16.692 (0.016) & 16.583 (0.014) & 16.344 (0.016) & 2 \\
2918.72   & \phn+69.9       &   18.470 (0.074) & 17.502 (0.029) & 16.767 (0.012) & 16.759 (0.010) & 16.457 (0.013) & 1 \\
2926.67   & \phn+77.9       &   18.797 (0.054) & 17.655 (0.034) & 16.963 (0.011) & 17.070 (0.012) & 16.688 (0.016) & 1 \\
2939.77   & \phn+91.0       &    \ldots        & 17.867 (0.018) & 17.419 (0.020) & 17.504 (0.018) & 17.224 (0.017) & 3 \\
2952.70   & +103.9          &  \ldots          & 18.055 (0.029) & 17.691 (0.013) & 17.850 (0.016) & 17.364 (0.019) & 2 \\
2967.70   & +118.9          &  \ldots          & 18.373 (0.018) & 18.076 (0.014) & 18.257 (0.015) & 17.731 (0.024) & 2 \\
2996.60   & +147.8          &  \ldots          & \ldots         &  \ldots        &  \ldots        & 18.253 (0.033) & 2 \\
3000.50   & +151.7          &  \ldots          & \ldots         & 18.712 (0.022) & 18.941 (0.017) & 18.361 (0.029) & 2 \\
3001.60   & +152.8          &  \ldots          & 18.934 (0.029) &     \ldots     &    \ldots      &  \ldots        & 2 \\
3008.70   & +159.9          &   \ldots         & \ldots         & 19.225 (0.035) & 19.431 (0.058) & 18.615 (0.061) & 4 \\
3010.70   & +161.9          &   \ldots         & 19.393 (0.076) & 19.173 (0.065) & 19.418 (0.025) & 18.607 (0.027) & 4 \\
3019.60   & +170.8          &  \ldots          & 19.307 (0.038) & 19.147 (0.019) & 19.311 (0.022) & 18.830 (0.040) & 2 \\
3030.57   & +181.8          &   \ldots         & 19.685 (0.036) & 19.460 (0.047) & 19.657 (0.038) & 19.188 (0.041) & 3 \\
3031.56   & +182.8          &   20.548 (0.145) & 19.975 (0.067) & 19.474 (0.046) & 19.724 (0.043) & 18.868 (0.055) & 1 \\
3034.54   & +185.7          &      \ldots      & 20.143 (0.090) & 19.432 (0.058) & 19.946 (0.078) & 18.968 (0.071) & 1 \\
3044.60   & +195.8          &   \ldots         & 19.665 (0.069) & 19.893 (0.037) & 20.107 (0.047) & 19.228 (0.041) & 4 \\
3045.50   & +196.8          &   \ldots         & 19.767 (0.059) & 19.798 (0.037) & 20.248 (0.105) & 19.223 (0.043) & 4 \\
3046.60   & +197.8          &   \ldots         & 19.694 (0.054) & 19.926 (0.039) & 20.057 (0.061) & 19.264 (0.083) & 4 \\
3050.52   & +201.7          &      \ldots      & 20.354 (0.138) & 19.855 (0.044) & 20.171 (0.083) & 19.148 (0.088) & 1 \\
3051.55   & +202.7          &      \ldots      & 19.904 (0.405) & 19.839 (0.126) & 20.249 (0.267) & 19.045 (0.144) & 3 \\
3282.90   & +434.1          &   \ldots         & 22.823 (0.293) & 22.551 (0.473) & \ldots         & 20.003 (0.121) & 5 \\
3283.80   & +435.0          &   \ldots         &  \ldots        & \ldots         & 22.732 (0.344) & \ldots         & 5 \\
3284.80   & +436.0          &   \ldots         & 22.516 (0.295) & 22.789 (0.666) & \ldots         &  \ldots        & 5 \\
3342.80   & +494.0          &   \ldots         &  \ldots        & 22.911 (0.350) & \ldots         &  \ldots        & 5 \\
\enddata
\tablenotetext{a} {Julian Date {\em minus} 2,450,000.}
\tablenotetext{b} {Days since T($B_{max}$).}
\tablenotetext{c} {Telescope: 1 = CTIO 1.3-m; 2 = Las Campanas 1.0-m; 
3 = CTIO 0.9-m; 4 = Univ. Arizona 1.54-m; 5 = Univ. Arizona 2.3-m; 6 = Magellan \#2 (Clay) Telescope.}
\end{deluxetable}

\begin{deluxetable}{clcccccc}
\tablewidth{0pt}
\tablecolumns{8}
\tabletypesize{\scriptsize}
\tablecaption{Near-Infrared Photometry of SN~2003gs\label{yjhk_data}}
\tablehead{
\colhead{UT date} & 
\colhead{JD\tablenotemark{a}} &
\colhead{Epoch\tablenotemark{b}} & 
\colhead{$Y$} &
\colhead{$J$} &
\colhead{$H$} &
\colhead{$K$} &
\colhead{Telescope\tablenotemark{b}}
}
\startdata

    Jul. 30 &  2850.91 & \phn\phn+2.1 &   13.44 (0.02)   & 13.41 (0.01)   & 13.38 (0.01)   & 13.40 (0.02)   & 1 \\
    Aug. 01 &  2852.85 & \phn\phn+4.1 &   \ldots         & 13.55 (0.01)   & 13.45 (0.01)   & 13.38 (0.02)   & 2 \\
    Aug. 01 &  2852.89 & \phn\phn+4.1 &   13.51 (0.02)   & 13.57 (0.02)   & 13.43 (0.02)   & 13.47 (0.02)   & 1 \\
    Aug. 03 &  2854.86 & \phn\phn+6.1 &   13.52 (0.02)   & 13.81 (0.02)   & 13.43 (0.01)   & 13.48 (0.02)   & 1 \\
    Aug. 05 &  2856.79 & \phn\phn+8.0 &   \ldots         & 14.01 (0.02)   & 13.47 (0.01)   & 13.43 (0.02)   & 2 \\
    Aug. 05 &  2856.88 & \phn\phn+8.1 &   13.53 (0.02)   & 13.95 (0.02)   & 13.46 (0.02)   &   \ldots       & 1 \\
    Aug. 07 &  2858.87 & \phn+10.1 &   13.47 (0.02)   & 14.07 (0.02)   & 13.48 (0.01)   & 13.46 (0.02)   &  1 \\
    Aug. 10 &  2861.90 & \phn+13.1 &   13.38 (0.02)   & 14.13 (0.02)   & 13.54 (0.01)   & 13.55 (0.02)   &  1 \\
    Aug. 11 &  2863.79 & \phn+15.0 &   \ldots         & 14.07  (0.02)  & 13.67 (0.01)   & 13.55 (0.02)   &  2 \\
    Aug. 13 &  2864.88 & \phn+16.1 &   13.32 (0.02)   & 14.09 (0.03)   & 13.65 (0.02)   & 13.68 (0.02)   &  1 \\
    Aug. 16 &  2867.81 & \phn+19.0 &   13.43 (0.02)   & 14.30 (0.02)   & 13.90 (0.02)   & 13.94 (0.03)   &  1 \\
    Aug. 20 &  2871.80 & \phn+23.0 &   \ldots         & 14.79  (0.03)  & 14.32 (0.01)   & 14.37 (0.02)   &  2 \\
    Aug. 21 &  2872.76 & \phn+24.0 &   13.91 (0.03)   & 15.06 (0.04)   & 14.40 (0.04)   & 14.49 (0.09)   &  1 \\
    Aug. 25 &  2876.83 & \phn+28.0 &   14.10 (0.02)   & 15.44 (0.03)   & 14.59 (0.03)   & 14.77 (0.04)   &  1 \\
    Aug. 31 &  2882.77 & \phn+34.0 &   14.51 (0.03)   & 15.96 (0.07)   & 15.02 (0.04)   & 15.01 (0.08)   &  1 \\
    Sep. 06 &  2888.88 & \phn+40.1 &   \ldots         & 16.55  (0.04)  & 15.31 (0.03)   & 15.47 (0.03)   &  2 \\
    Sep. 08 &  2890.77 & \phn+42.0 &   14.93 (0.03)   & 16.79 (0.12)   & 15.25 (0.05)   & 15.54 (0.06)   &  1 \\
    Sep. 15 &  2897.80 & \phn+49.0 &      \ldots      & 16.98 (0.11)   & 15.41 (0.05)   & \ldots         &  1 \\
    Sep. 18 &  2900.78 & \phn+52.0 &      \ldots      & 17.36  (0.05)  & 15.86 (0.06)   & 16.13 (0.05)   &  2 \\
    Sep. 29 &  2911.75 & \phn+63.0 &      \ldots      & 17.69 (0.22)   & 15.94 (0.07)   & \ldots         &  1 \\
    Oct. 06 &  2918.74 & \phn+70.0 &      \ldots      & 18.12 (0.20)   & 16.12 (0.09)   & \ldots         &  1 \\
    Oct. 14 &  2926.69 & \phn+77.9 &      \ldots      & 18.51 (0.28)   & 16.60 (0.08)   & \ldots         &  1 \\
2004 Aug. 03 & 3220.82 & +372.0    &      \ldots      & $>$21.50       & $>$20.60       & $>$20.80       &  3 \\

%
%
\enddata

\tablenotetext{a} {Julian Date minus 2,450,000.}
\tablenotetext{b} {1 = CTIO 1.3-m + ANDICAM.  2 = ESO 3.6-m NTT + SofI.
3 = VLT + ISAAC.}
\end{deluxetable}

\begin{deluxetable}{cccc}
\tablewidth{0pt}
\tablecolumns{4}
\tablecaption{S-corrections for CTIO 1.3-m near-IR photometry\tablenotemark{a}\label{irscorr}}
\tablehead{
\colhead{$t$} &
\colhead{$\Delta J$} &
\colhead{$\Delta H$} &
\colhead{$\Delta K$} 
}
\startdata
2.1  &   [0.028]   & [$-$0.040] &   [0.031]  \\
4.1  &    0.028    &  $-$0.053  &    0.031   \\
6.1  &    0.009    &  $-$0.063  &    0.019   \\
8.1  &    0.000    &  $-$0.064  &  \ldots    \\
10.1 &    0.004    &  $-$0.050  &    0.022   \\
13.1 &    0.011    &  $-$0.027  &    0.027   \\
16.1 & $-$0.027    &  $-$0.026  & $-$0.019   \\
19.0 & $-$0.043    &  $-$0.022  & $-$0.032   \\
24.0 & $-$0.063    &  $-$0.014  & $-$0.045   \\
28.0 & $-$0.081    &  $-$0.012  & $-$0.057   \\
34.0 & $-$0.082    &  $-$0.014  & $-$0.050   \\
42.0 & $-$0.076    &  $-$0.019  & $-$0.034   \\
49.0 & $-$0.059    &  $-$0.025  & \ldots     \\
63.0 & $-$0.020    &  $-$0.037  & \ldots     \\
70.0 &   [0.000]   & [$-$0.037] & \ldots     \\
77.9 &   [0.000]   & [$-$0.037] & \ldots     \\
\enddata
\tablenotetext{a}{
These values are to be {\em added} to the values in Table \ref{yjhk_data}
to place the ANDICAM near-IR photometry on the photometric system of \citet{Per_etal98}.  
$t$ is the number of days since $B$-band maximum.}
\end{deluxetable}

\begin{deluxetable}{ccccc}
\tablewidth{0pt}
\tablecolumns{5}
\tablecaption{Low order polynomial fits to $V-H$ and $V-K$ colors\tablenotemark{a}\label{vmircolors}}
\tablehead{
\colhead{Color Index:} &
\colhead{$V-H$} &
\colhead{$V-H$} &
\colhead{$V-K$} &
\colhead{$V-K$} 
 }
\startdata
   Range (d)       &  [$-$9,+16.5]  & [+18.5,+88.5]   & [$-$8,+16]  &  [18.5, 85.5] \\
    a$_0$           &    0.091       &    1.399        &   $-$0.152  &     1.671     \\
    a$_1$           & $-$0.5126819E$-$02 & $-$0.3641677E$-$01 & $-$0.2787401E$-$02 & $-$0.5979778E$-$01 \\
    a$_2$           &   +0.3556262E$-$02 &   +0.3332700E$-$03 &   +0.7684491E$-$02 &   +0.7580789E$-$03 \\
    a$_3$           & $-$0.1658202E$-$04 & $-$0.1711514E$-$05 & $-$0.2259151E$-$03 & $-$0.4665809E$-$05 \\
    rms (mag)       &  $\pm$0.172    &    $\pm$0.101   &  $\pm$0.209 &   $\pm$0.107   \\
 $\chi ^2 _{\nu}$   &   23.9         &     2.1         &  15.5       &     2.9    \\
\enddata
\tablenotetext{a}{
Fits are of the form: $V-H$ or $V-K$ = $a_0 + \Sigma (a_i t^i)$.}
\end{deluxetable}

\begin{deluxetable}{lcccc}
\tablewidth{0pt}
\tablecolumns{5}
\tablecaption{Near-IR absolute magnitudes at maximum\tablenotemark{a}\label{absmag_avgs}}
\tablehead{
\colhead{Group} &
\colhead{Filter} &
\colhead{$\langle$M$\rangle$} &
\colhead{$\sigma _x$} &
\colhead{N} 
}
\startdata
Peak early: &              &               &              &       \\
            &    $J$       &  $-$18.614    &  $\pm$0.158  &   25  \\
            &    $H$       &  $-$18.308    &  $\pm$0.149  &   25  \\
            &    $K$       &  $-$18.436    &  $\pm$0.153  &   23  \\
            &              &               &              &       \\
Peak late:  &              &               &              &       \\
            &    $J$       &  $-$17.802    &  $\pm$0.133  &   5  \\
            &    $H$       &  $-$17.867    &  $\pm$0.067  &   5  \\
            &    $K$       &  $-$17.792    &  $\pm$0.064  &   4  \\
\enddata
\tablenotetext{a}{
Type Ia SNe typically peak in the near-IR 3 days before
T($B_{max}$).  Those that peak in the near-IR a few days after T($B_{max}$)
are fast decliners at optical wavelengths and are faint in all bands.}
\end{deluxetable}

\begin{deluxetable}{ccc}
\tablewidth{0pt}
\tablecolumns{3}
\tablecaption{$^{56}$Ni masses estimated using Arnett's rule\tablenotemark{a}\label{nickel}}
\tablehead{
\colhead{SN} &
\colhead{\dmm} &
\colhead{$M({^{56}\mathrm{Ni}})/M_\odot$} 
}
\startdata
2003gs        &    1.83       & 0.25    \\
1999by        &    1.90       & 0.09     \\
2005bl        &    1.93       & 0.10     \\
1991bg        &    1.94       & 0.07      \\
1998de        &    1.95       & 0.06      \\
\enddata
\tablenotetext{a}{
Values of \dmm\, for all SNe other than SN~2003gs are taken from
               \citet[][and references therein]{Tau_etal08}.}
\end{deluxetable}

\clearpage

\figcaption[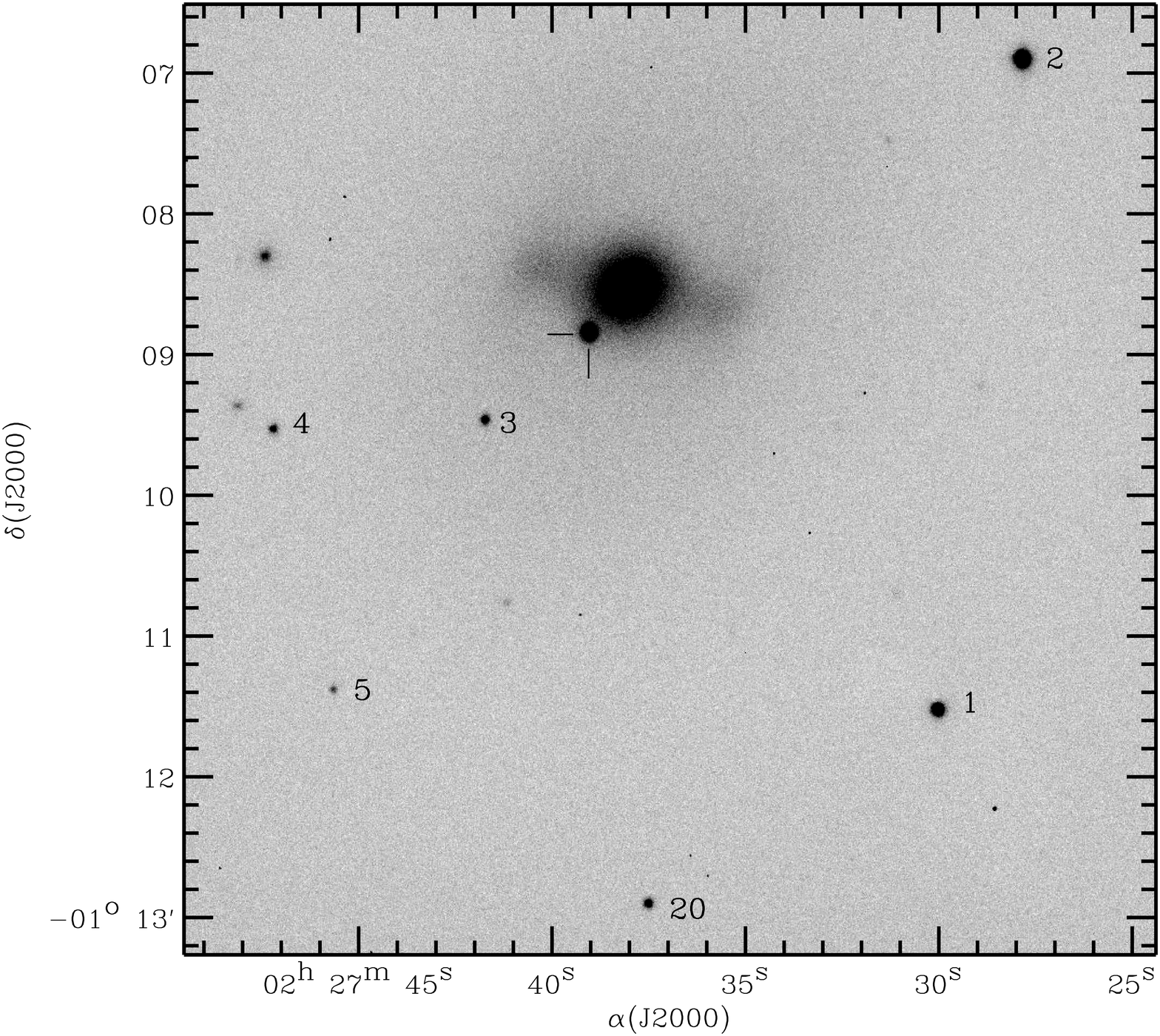] {Finder chart.
\label{finder}
}

\figcaption[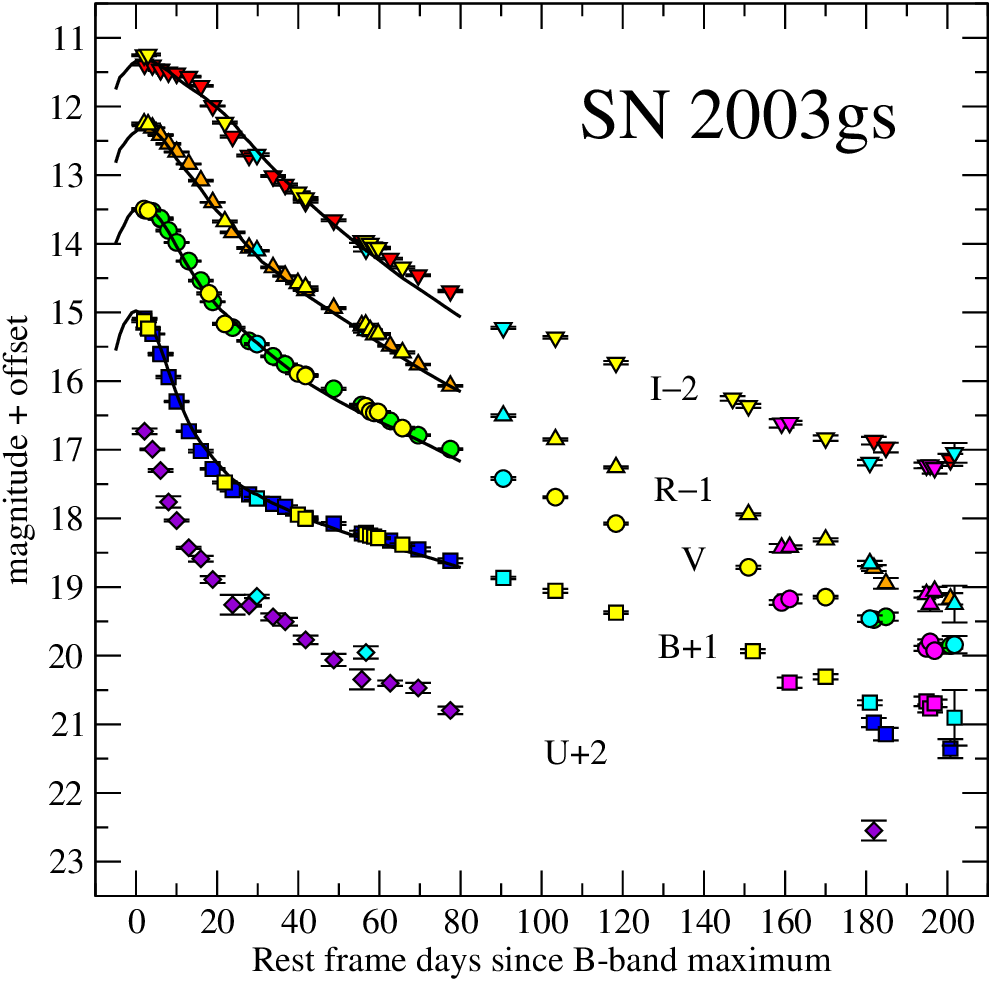] {Optical light curves of SN~2003gs.  
\label{03gs_ubvri}
}

\figcaption[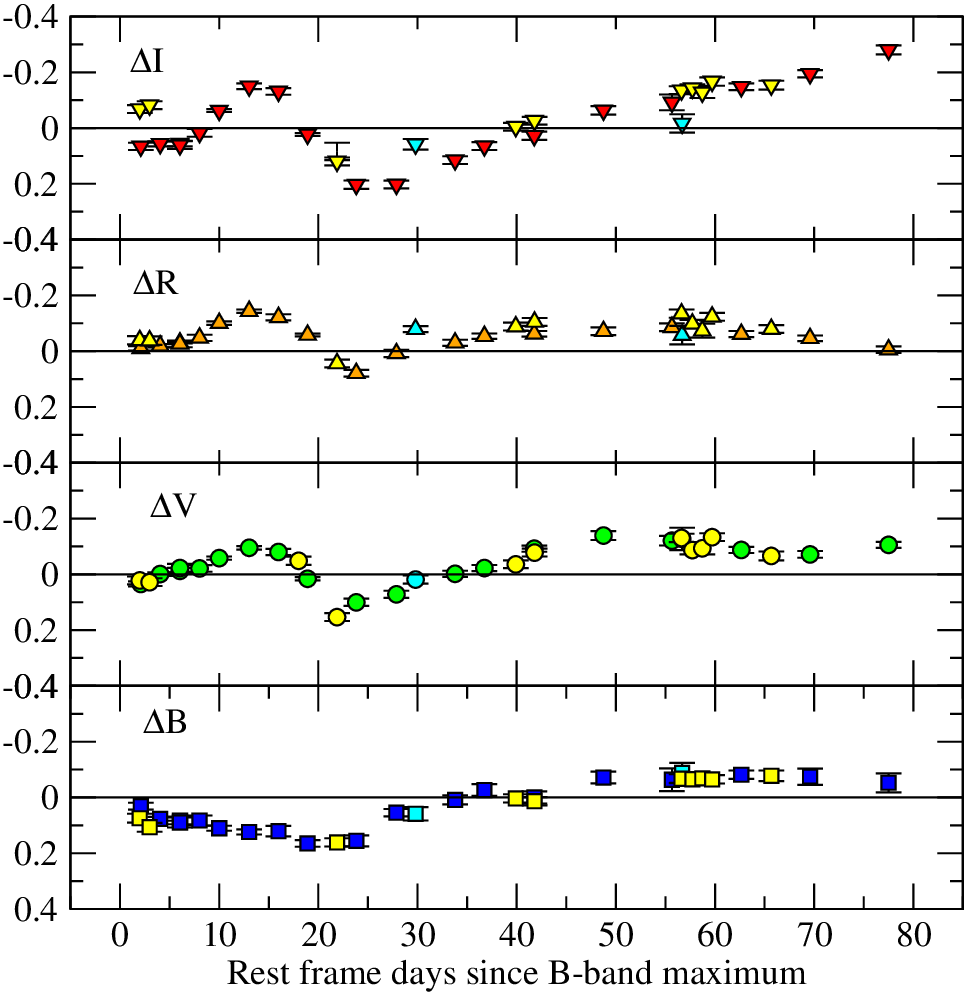] {
Residuals of $BVRI$ photometry with respect to \dmm\ = 1.83 templates
shown in Fig. \ref{03gs_ubvri}.
\label{resids}
}

\figcaption[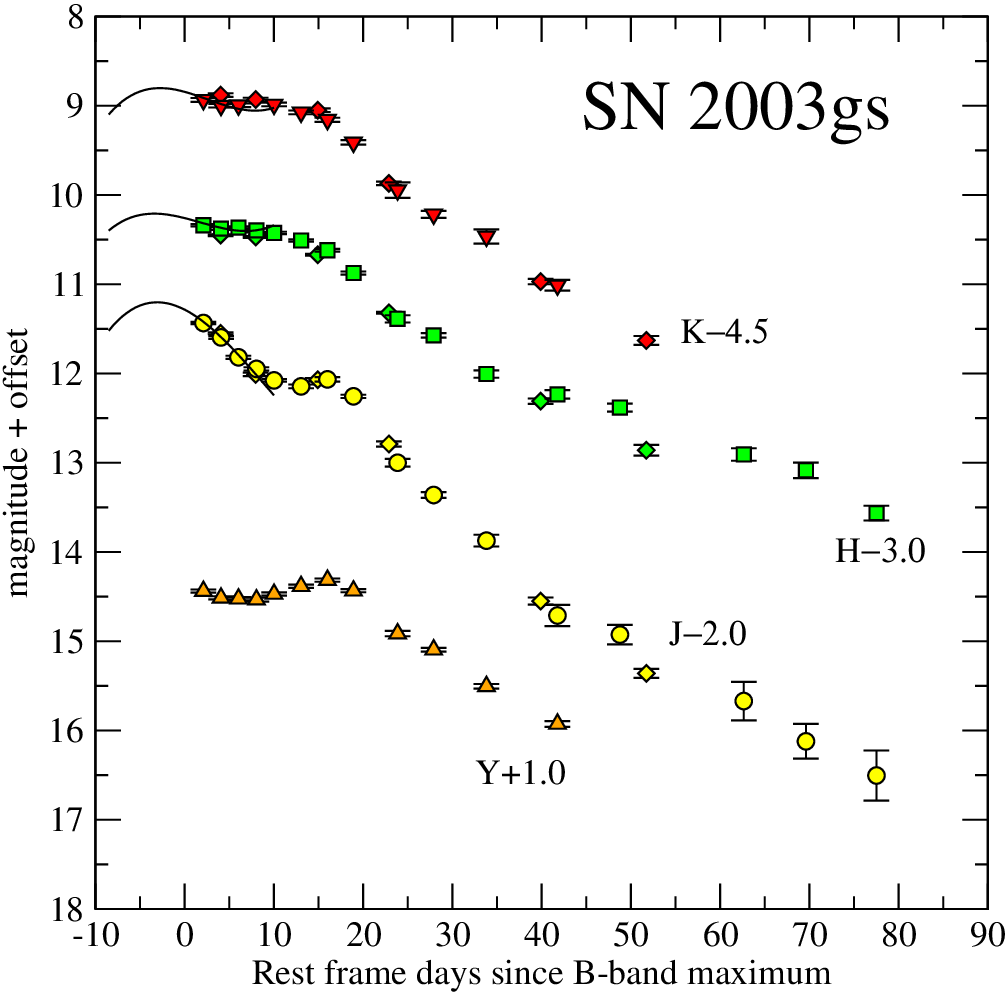] {$YJHK$ light curves of SN~2003gs.
\label{03gs_yjhk}
}

\figcaption[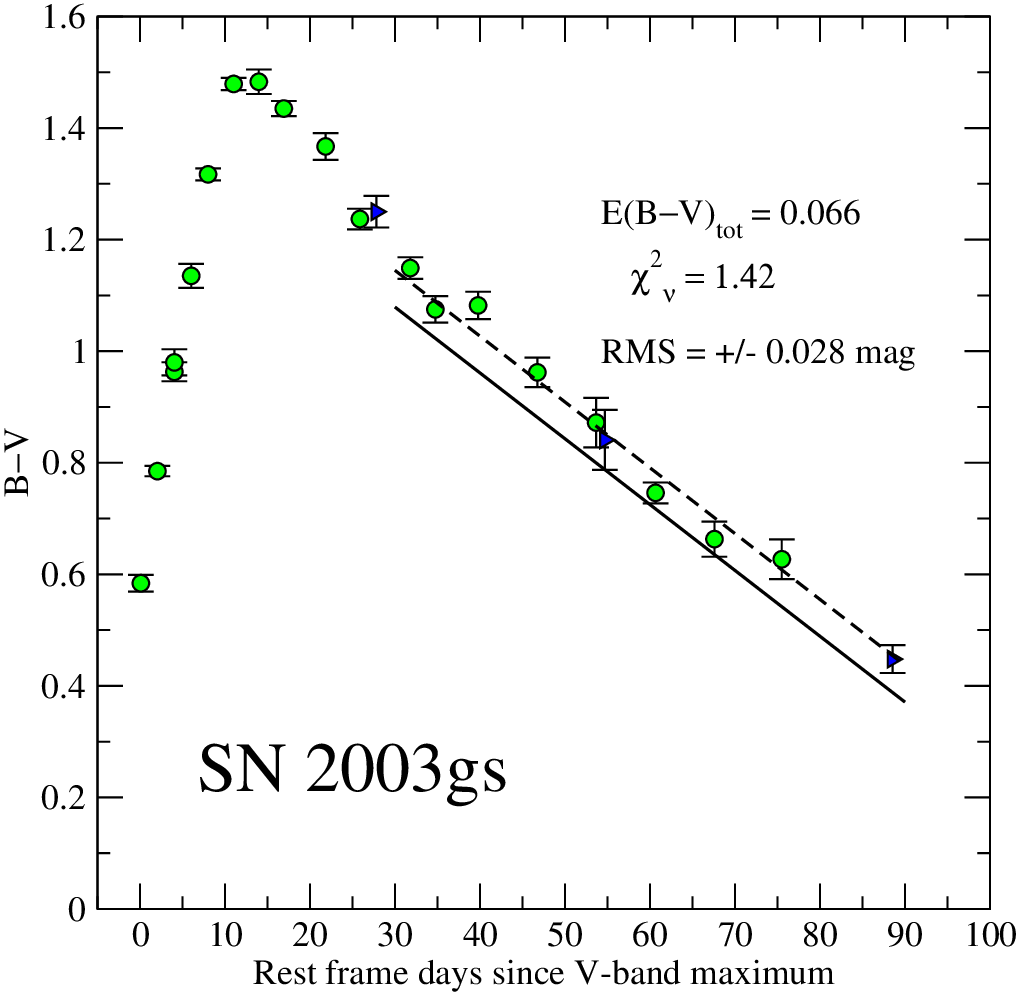] {$B-V$ color curve and E($B-V$).
\label{03gs_bv}
}

\figcaption[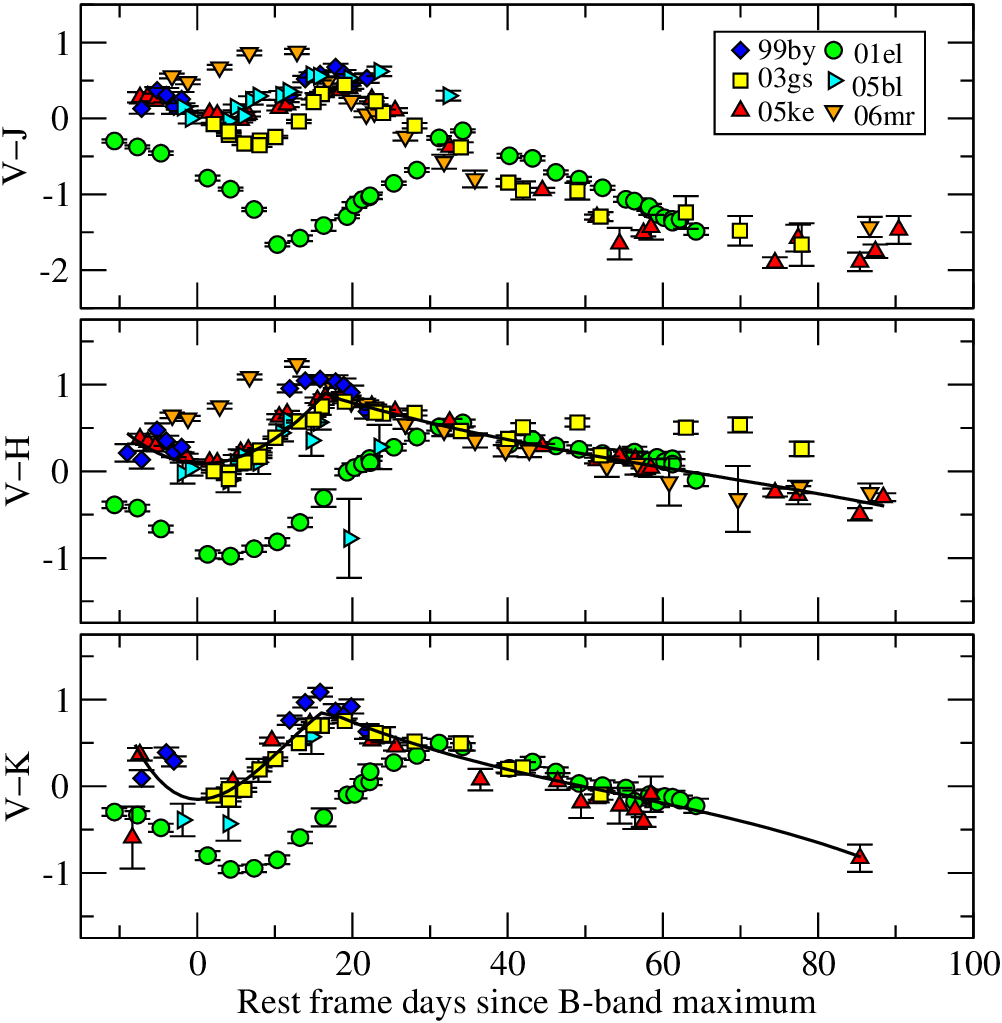] {$V$ minus IR color curves of fast decliners and
SN~2001el.
\label{vjhk}
}

\figcaption[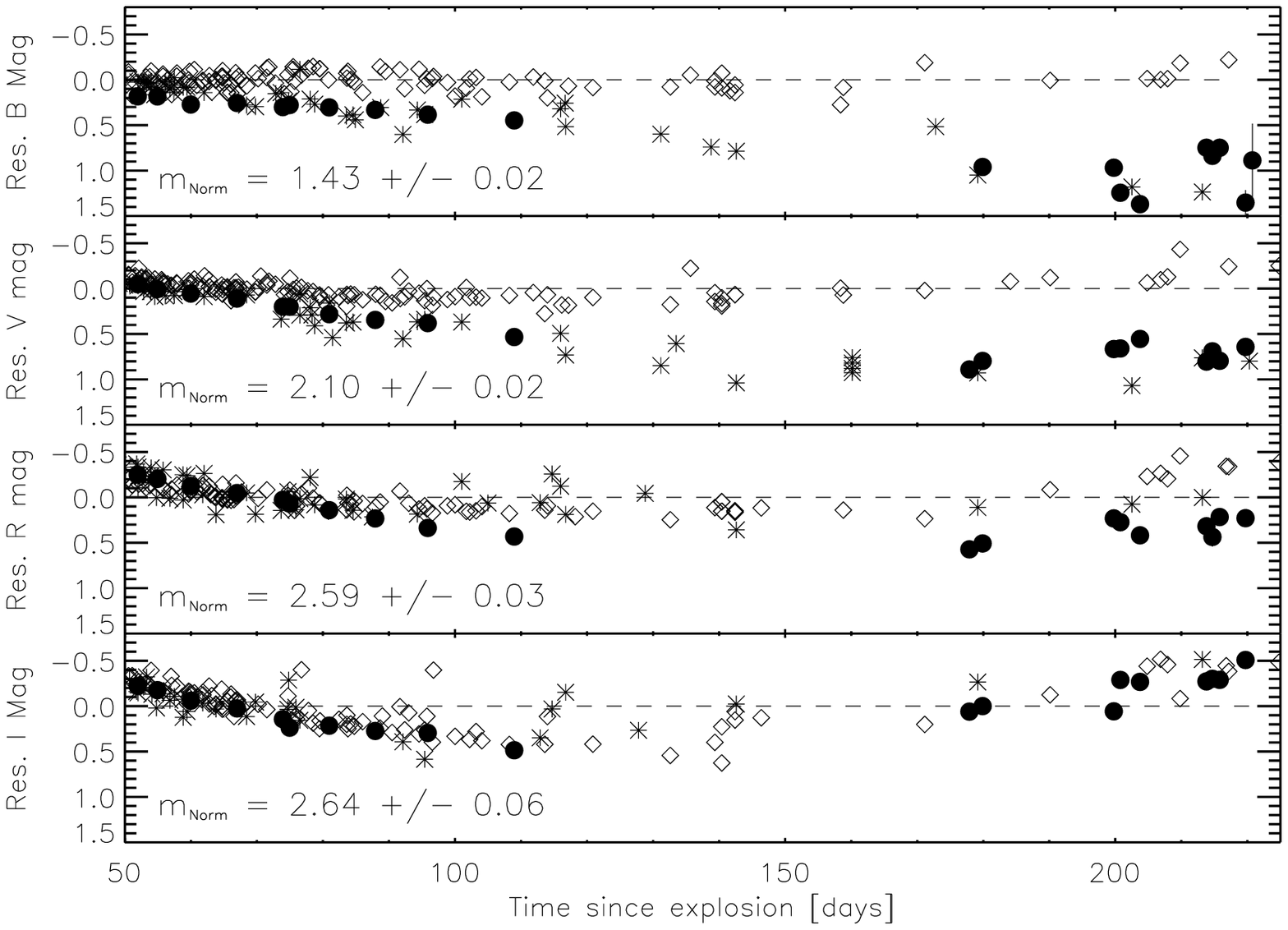] {Comparison of late-time light curves.
\label{late_time}
}

\figcaption[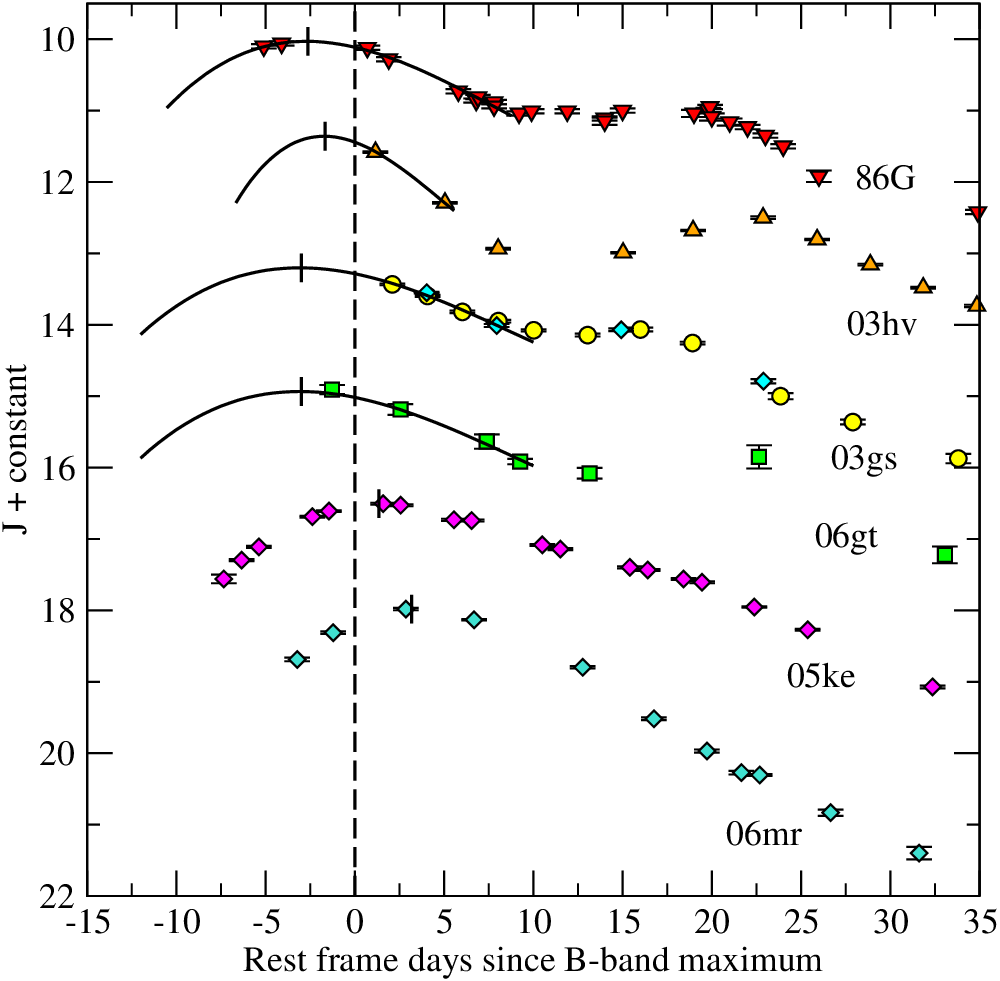] {Comparison of $J$-band light curves.
\label{jband}
}

\figcaption[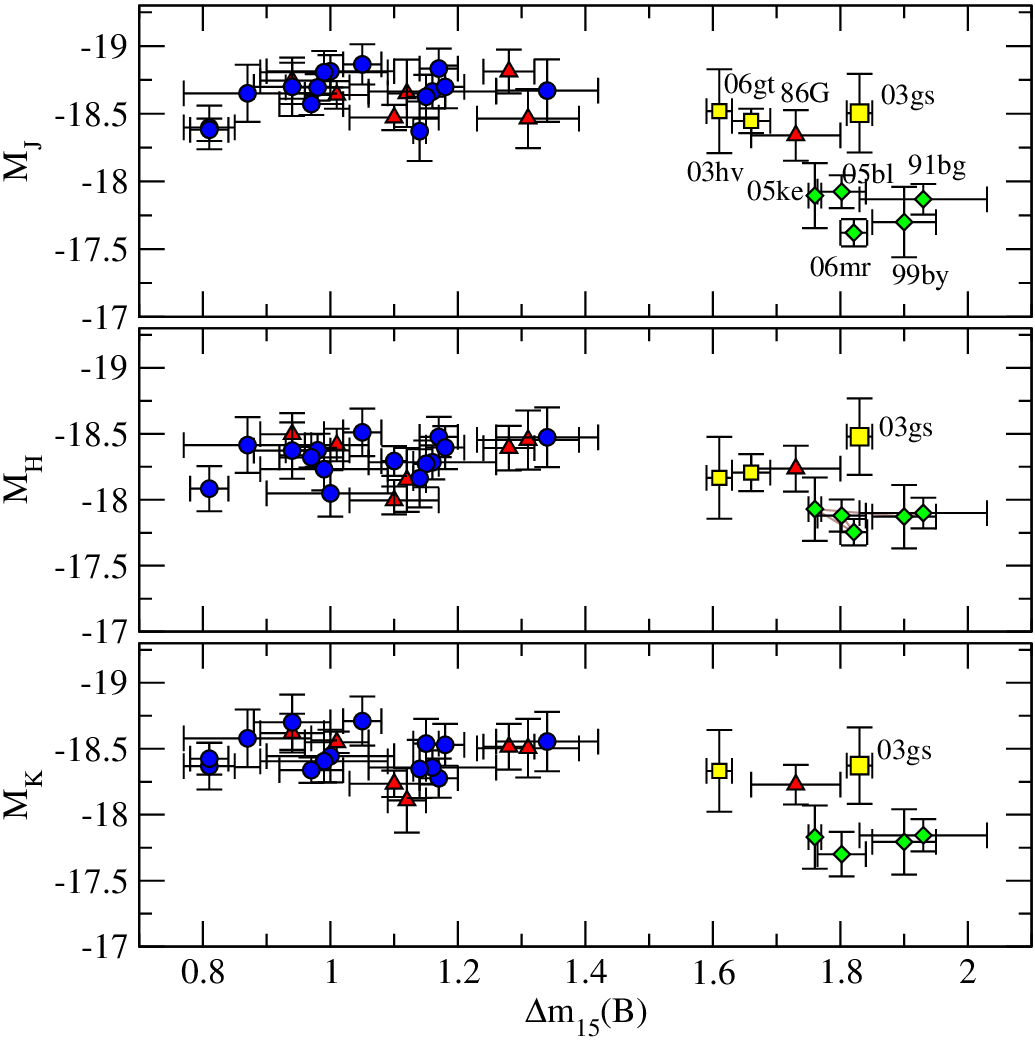] {Near-IR decline rate relations.
\label{absmags}
}

\figcaption[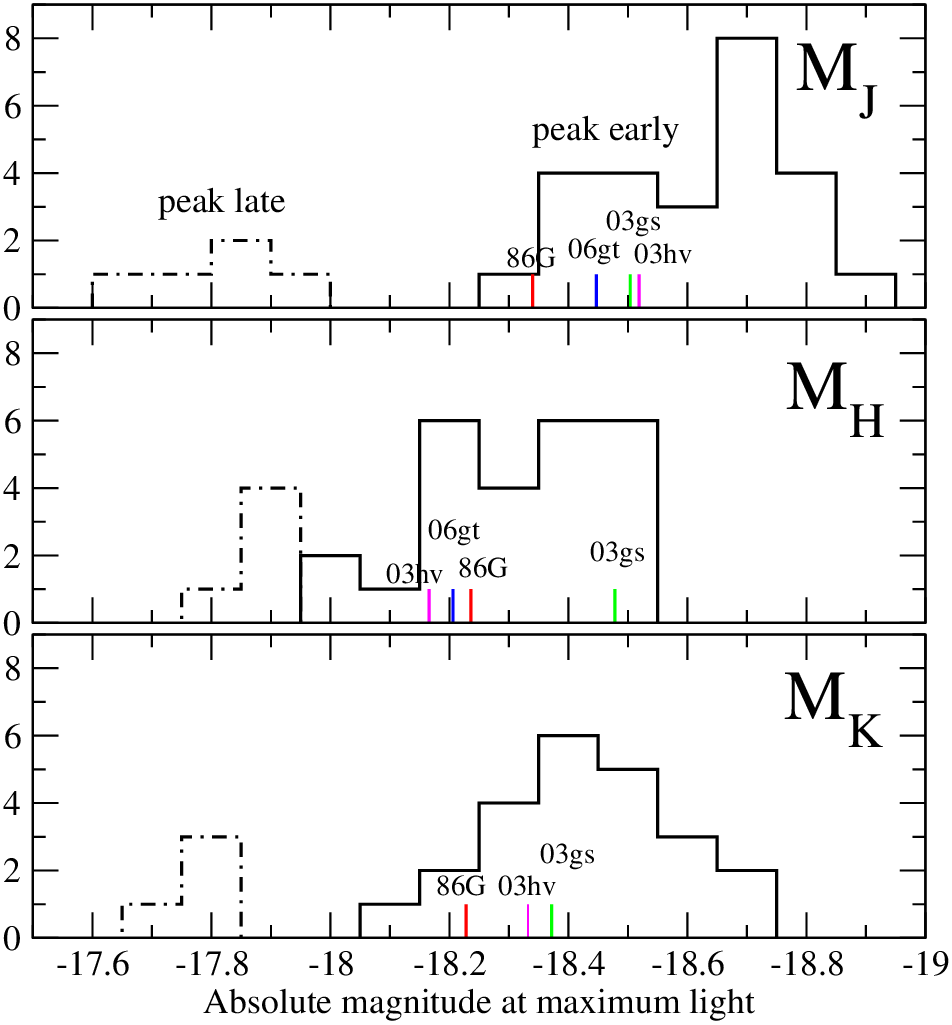] {Histograms of absolute magnitudes at
maximum light.
\label{early_late}
}

\figcaption[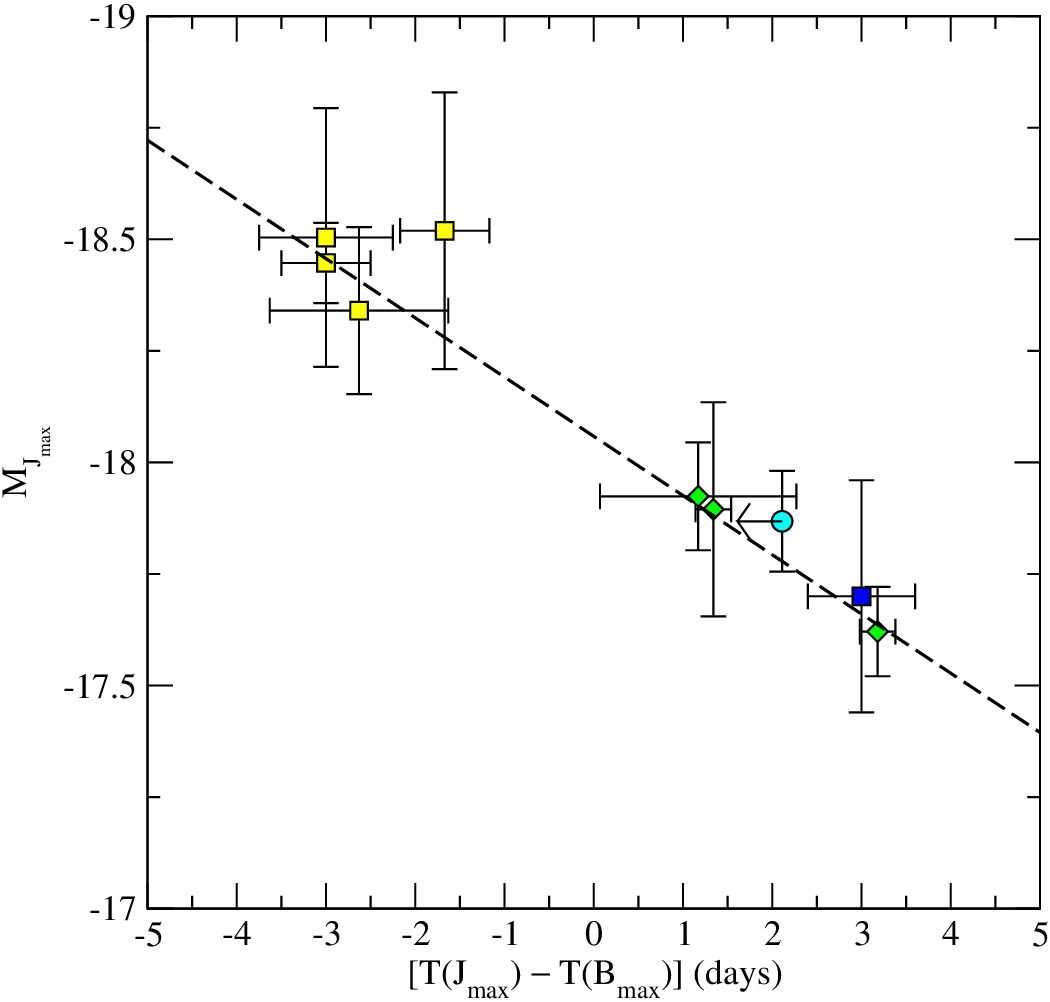] {$J$-band absolute magnitudes vs. relative
time of maximum with respect to $B$-band.
\label{delay}
}

\figcaption[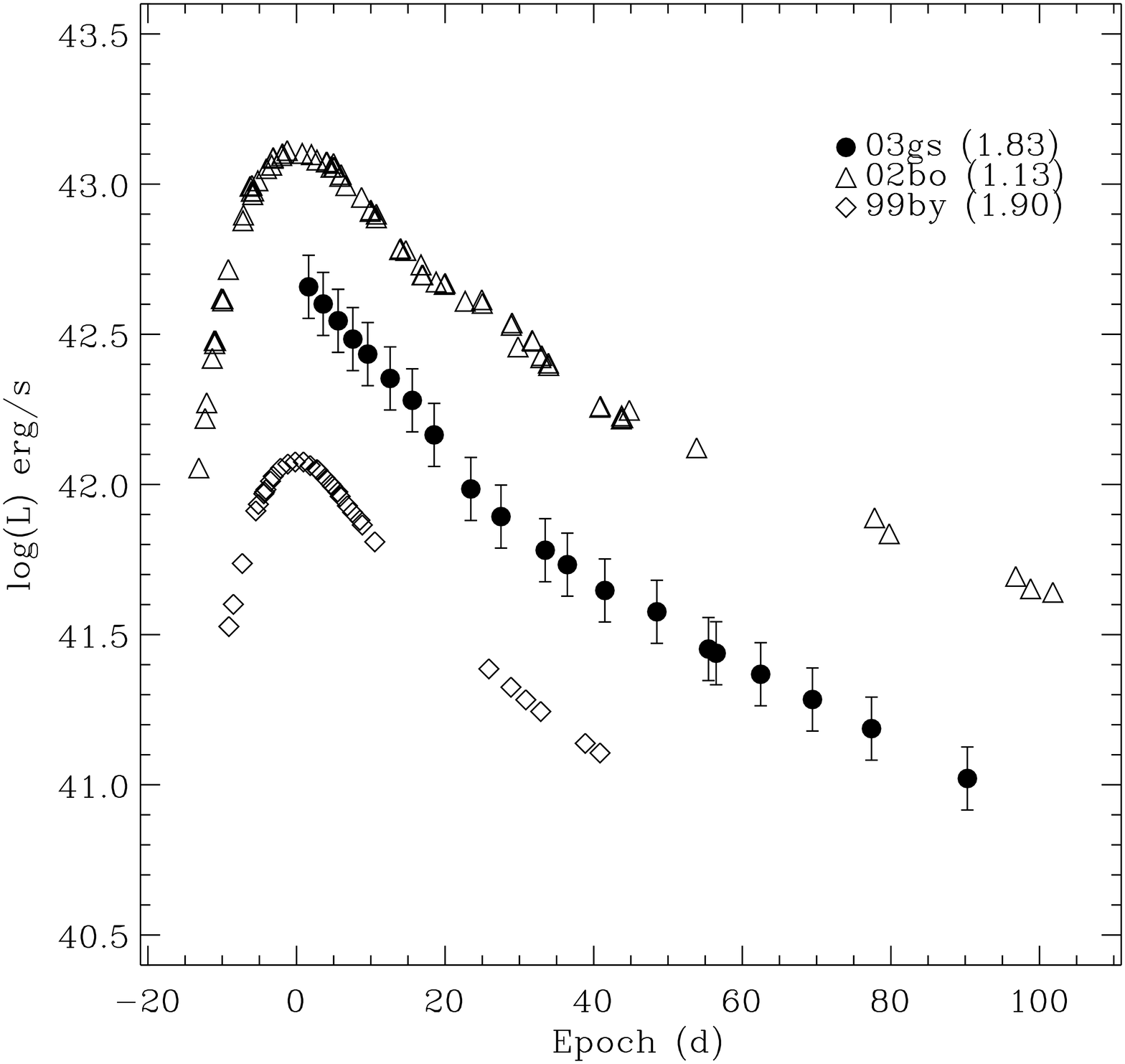] {Bolometric light curves.
\label{blc_jensen}
}

\figcaption[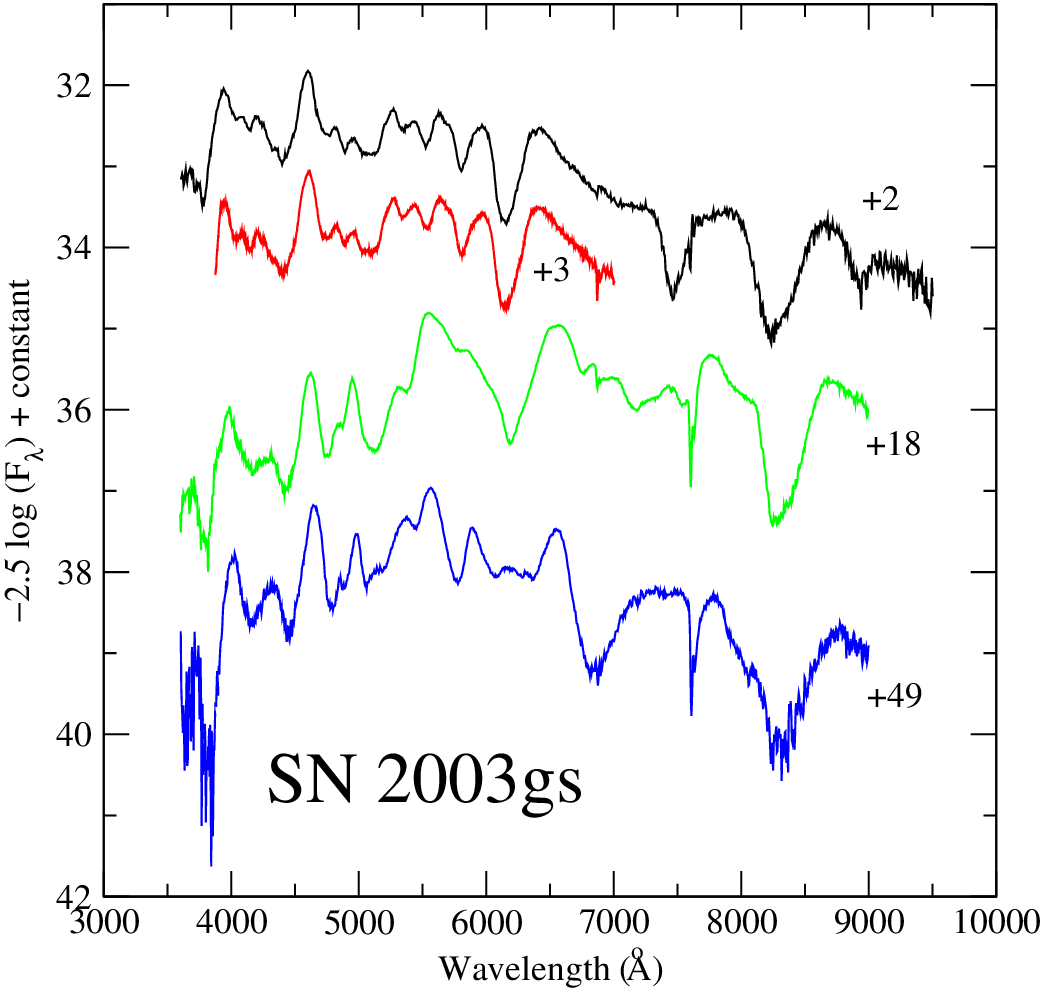] {Four optical spectra of SN~2003gs.
\label{spectra}
}

\clearpage

\begin{figure}
\plotone{chart.eps}
{Fig. \ref{finder}.
Finder chart for NGC 936, SN~2003gs, and some
field stars in our Galaxy.  This 6.3 $\times$ 6.3 arcmin image is a 
20 sec $V$-band exposure obtained at 08:56 UT on 1 August 2003 ($t\sim$\,4\,d) with 
the CTIO 1.3-m telescope.  North is up and east to the left.  The SN
is located 13.4 arcsec east and 14.6 arcsec south of the galaxy core.
}
\end{figure}

\begin{figure}
\plotone{03gs_ubvri.eps}
{Fig. \ref{03gs_ubvri}.
Optical light curves of SN~2003gs.  Yellow symbols 
represent data obtained at Las Campanas.  Cyan symbols represent data 
obtained with the CTIO 0.9-m.  Magenta symbols represent data obtained with 
the University of Arizona 1.54-m telescope.  All other data were obtained with the
CTIO 1.3-m telescope.  The data have been shifted vertically
for clarity by the amounts indicated.  The solid lines represent the \dmm\ = 1.83
set of light curve templates of \citet{Pri_etal06}, adjusted to minimize the total
$\chi ^2$ of the fits.
}
\end{figure}

\begin{figure}
\plotone{resids.eps}
{Fig. \ref{resids}.
Residuals of $BVRI$ photometry with respect to \dmm\ = 1.83 templates
shown in Fig. \ref{03gs_ubvri}.
}
\end{figure}

\begin{figure}
\plotone{03gs_yjhk.eps}
{Fig. \ref{03gs_yjhk}.
Near-IR photometry.  The diamond-shape symbols represent data
obtained with the NTT.  The other data were taken with the CTIO 1.3-m.
The data have been offset vertically for clarity by the number of magnitudes
given. We have added the $JHK$ maximum light templates of \citet{Kri_etal04b}.
}
\end{figure}

\begin{figure}
\plotone{03gs_bv.eps}
{Fig. \ref{03gs_bv}.
$B-V$ colors of SN~2003gs as a function of time since $V$-band
maximum.  The right pointing triangles are CTIO 0.9-m data.
The other data are from the CTIO 1.3-m, with the S-corrections applied.
The solid line is the locus of \citet{Lir95} for unreddened Type
Ia SNe.  The dashed line represents an offset corresponding to
E($B-V$) = 0.066 mag.
}
\end{figure}

\begin{figure}
\plotone{vjhk.eps}
{Fig. \ref{vjhk}.
$V-J$, $V-H$, and $V-K$ colors of Type Ia SNe.
The photometry has been corrected for Galactic reddening
and host galaxy reddening.  The low-order polynomial fits to
subsets of the $V-H$ and $V-K$ data given in Table \ref{vmircolors}
are also shown here.  Prior to $t$ = 17 d the fits have a
large reduced $\chi^2$, but after that the nearly linear fits
are considerably tighter and even match the colors of the 
normal Type Ia SN 2001el.
}
\end{figure}

\begin{figure}
\plotone{late_time.ps}
{Fig. \ref{late_time}.
From 50 to 200 days after the time of explosion Type Ia SNe
light curves exhibit a reasonably linear decline.  Here we plot the
differences of the photometry with respect to the $BVRI$ rates of
decline of normal objects.  Symbols: diamonds = normal Type Ia SNe;
dots = SN~2003gs; asterisks = other fast decliners.
}
\end{figure}

\begin{figure}
\plotone{jband.eps}
{Fig. \ref{jband}.
$J$-band photometry of six fast declining Type Ia SNe.  The data of SN~2003hv
have been offset by $-$1.5 mag; 2006gt by $-$3.0 mag; 2005ke by +2.5 mag; 
and 2006mr by +4.0 mag.  The peaks of the top four objects are fitted
with the stretchable template of \citet{Kri_etal04b}.  The stretch factors
were 0.877 for SN~1986G, 0.556 for 2003hv, 1.00 for 2003gs, and 1.00 for 2006gt.
The top four objects peaked prior to T($B_{max}$), while the bottom two
peaked after T($B_{max}$).  Note also the weak secondary maxima in the bottom
two objects. 
}
\end{figure}

\begin{figure}
\plotone{absmag14.eps}
{Fig. \ref{absmags}.
Near-IR decline rate relations for Type Ia SNe (absolute
magnitudes at peak brightness vs. \dmm).  Blue dots represent objects
from \citet{Kri_etal04c} which are in the smooth Hubble flow, with
radial velocities in the CMB frame greater than 3000 km s$^{-1}$.
Red triangles represent objects from \citet{Kri_etal04c} which are
not far enough to be in the smooth Hubble flow; their distances were
determined via Cepheids or SBF distances.  SNe 2005bl and 2006gt are the only
fast decliners that have Hubble flow distances.  The objects
that peak $\sim$3 days before $B$-band maximum constitute a brighter sub-sample
in each sub-diagram.  The objects that peak a few to several days after
$B$-band maximum are all faint by comparison.  
}
\end{figure}

\begin{figure}
\plotone{early_late.eps}
{Fig. \ref{early_late}.
Histograms of the $J$-, $H$-, and $K$-band absolute magnitudes
of Type Ia SNe plotted in Fig. \ref{absmags}.  We have grouped the
objects according to the relative times of their infrared peaks compared
to the time of $B$-band maximum.  Those that peak early, including the
fast decliners SNe 1986G, 2003gs, and 2006gt are brighter at maximum light than
fast decliners that peak several days after T($B_{max}$).
}
\end{figure}

\begin{figure}
\plotone{delay.eps}
{Fig. \ref{delay}.
The $J$-band absolute magnitudes at maximum light for nine fast declining
Type Ia SNe vs. the relative time of the $J$-band and $B$-band maxima.
The dashed line is a fit to the points representing SNe 
2003gs, 2006gt, 1986G, and 2003hv (upper left), 2005bl, 2005ke, and 2006mr
(green diamonds, lower right).  The cyan circle is for SN~1991bg and is
an upper limit in the time domain.  The blue square is for SN~1999by
and assumes that the time of $J$-band maximum occurred at the same time
as the $I$-band maximum.
}
\end{figure}

\begin{figure}
\plotone{blc_jensen.eps}
{Fig. \ref{blc_jensen}.
Quasi-bolometric light curves for SN~2003gs,
SN~2002bo, and SN~1999by; \dmm\ for each SN is indicated
in parentheses. The error bars include uncertainties in
distance and reddening, and are shown for SN~2003gs only.
}
\end{figure}

\begin{figure}
\plotone{spectra.eps}
{Fig. \ref{spectra}.
Four optical spectra of SN~2003gs.  The numbers of days since
T($B_{max}$) are given in the plot.
}
\end{figure}

\end{document}